\documentstyle[epsf,rotate]{mn}

\def\bj{\hbox{$B_J$}}
\def\Bj{\hbox{$B_J$}} 
\def\b{\hbox{$B$}}

\def\r{\hbox{$R$}}
\def\R{\hbox{$R$}}
\def\i{\hbox{$I$}}

\def\Kn{\hbox{$Kn$}}
\def\kn{\hbox{$Kn$}}
\def\Sr{{\rm\thinspace Sr}}
\def\Jy{{\rm\thinspace Jy}}

\def\MHz{{\rm\thinspace MHz}}
\def\GHz{{\rm\thinspace GHz}}

\def\refs{\par \noindent \hang}
\def\ra{{\rm{RA}}}
\def\dec{{\rm{Dec}}}
\def\d{\hbox{$^\circ$}}
\def\h{\hbox{$^{\rm h}$}}
\def\ax{\approx}
\def\one_wide{9cm}
\def\two_wide{18cm}

\title{The Parkes Half-Jansky Flat-Spectrum Sample}
\author[M.J. Drinkwater et al.]
{M. J. Drinkwater$^1$,  
R. L. Webster$^2$, P. J. Francis$^2$, 
J. J. Condon$^3$, S. L. Ellison$^{4,1}$,\cr
D. L. Jauncey$^5$,
J. Lovell$^6$, B. A. Peterson$^7$,
A. Savage$^1$\thanks{Present address: ``Holly Farm'',
Binnaway Road, Coonabarabran, New South Wales 2357, Australia.}\\
$^1$Anglo-Australian Observatory, Coonabarabran, New South Wales 2357, Australia\\
$^2$School of Physics, University of Melbourne, Parkville, Victoria 3052, Australia\\
$^3$National Radio Astronomy Observatory\thanks{The National Radio Astronomy Observatory is operated by
Associated Universities, Inc., under cooperative agreement
with the National Science Foundation.}, 
520 Edgemont Road, 
Charlottesville, VA 22903, USA\\
$^4$Physics Department, University of Kent at Canterbury, Canterbury, Kent CT2 7NP, England\\
$^5$Australia Telescope National Facility, P.O. Box 76, Epping, New South Wales 2121, Australia\\
$^6$Physics Department, University of Tasmania, GPO Box 252C, Hobart, Tasmania 7001, Australia\\
$^7$Mt. Stromlo \& Siding Spring Observatories, Australian National University,
Private Bag, Weston Creek, A.C.T. 2611, Australia\\
}
\date{This is a preprint of a paper accepted for publication in MNRAS.}

\begin{document}

\maketitle



\begin{abstract}

We present a new sample of Parkes half-Jansky flat-spectrum radio
sources having made a particular effort to find any previously
unidentified sources. The sample contains 323 sources selected
according to
a flux limit of 0.5\Jy\ at 2.7\GHz, 
a spectral index measured between 2.7 and 5.0\GHz\ of
$\alpha_{2.7/5.0} > -0.5$, where $S(\nu) \propto \nu^{\alpha}$,
Galactic latitude $|b|>20\d$ and 
$-45\d<$ Declination (B1950) $<+10\d $. 
The sample was selected from a region 3.90 steradians in area.

We have obtained accurate radio positions for all the unresolved
sources in this sample and combined these with accurate optical
positions from digitised photographic sky survey data to check all the
optical identifications. We report new identifications based on \R-
and \Kn-band imaging and new spectroscopic measurements of many of the
sources. We present a catalogue of the 323 sources of which 321 now have
identified optical counterparts and 277 have measured spectral
redshifts.

\end{abstract}


\section{Introduction}

The Southern sky was surveyed at 2.7\GHz\ by the Parkes radio
telescope between 1968 and 1979 (see Bolton, Savage \& Wright 1979,
and references therein), resulting in a catalogue of more than 10,000
radio sources. Over this period, an extensive programme of optical
identifications was undertaken. In its early stages, this programme
was frustrated by lack of a Southern radio calibrator grid, poor radio
positions (the original Parkes positions were only accurate to
10--15~arcsec) and a lack of optical sky survey plates. Modern
methods, using accurate (better than 1 arcsec) radio positions and
complete catalogues of digitised optical sky survey data, with the
radio and optical reference frames tied to an accuracy of better than
100 milliarcsec (Johnston et al.\ 1995) now allow unambiguous
optical identification of most of the radio sources, supplemented by
CCD imaging for the remainder. In this paper we present
new identifications of a sample of Parkes flat-spectrum radio sources
using these techniques. The Parkes Catalogue contains both steep- and
flat-spectrum sources.  Radio samples are biased towards
core-dominated quasars if the flat-spectrum sources are selected and
towards lobe-dominated quasars and galaxies for steep-spectrum
sources.  Since the scientific questions of primary interest to us are
related to core-dominated quasars, we concentrate on flat-spectrum
sources in this study.

Other workers have compiled a number of complete samples of radio
sources. Each has been selected on different criteria, leading to the
inclusion of different objects. Low frequency samples contain more
radio galaxies than quasars, high frequency (i.e. flat-spectrum)
samples reverse that bias, and lower flux limits increase the mean
redshift of the objects in the sample.  Four notable samples are the
3CR Sample (Spinrad et al.\ 1985, Laing, Riley \& Longair 1983), the 2
Jansky Sample (Wall \& Peacock 1985), the 1 Jansky Sample (K\"uhr et
al.\ 1981; Stickel, Meisenheimer \& K\"uhr 1994), and the Parkes
Selected Regions (Dunlop et al.\ 1989).

The 3CR sample 
comprises 173 sources selected with 
$S_{178\MHz} > 10$\Jy\ over an area of 4.23\Sr.  
The high flux limit biases this sample towards lower
redshift objects (18\% have $z>1$), and the use of a low frequency
biases the sample towards steep-spectrum radio galaxies.    
The 2 Jansky
sample, which was selected at 2.7\GHz\ over an area of 9.81\Sr, 
contains 233 objects which are
mainly steep-spectrum, again biasing the sample towards low-redshift
radio galaxies.
The 1 Jansky
sample was selected at 5\GHz, also over 9.81\Sr\ of sky and
comprises 518 sources.  55\% are flat-spectrum sources, 
and of those $\sim 90\%$ are
quasars or BL Lacs; many are Parkes sources.  
Finally, the Parkes Selected Regions (a total of 0.075\Sr)
contain 178
sources with $S_{2.7\GHz} > 0.1\Jy$, most of which are
steep-spectrum extended sources identified as galaxies. Only 23\% are
flat-spectrum sources, but these objects have a distribution of
properties similar to our sample.

Our primary interest in this paper is the compilation of a large
unbiased sample of radio-selected quasars. Note that we use the standard
definitions of ``quasar'' for radio-loud sources and quasi-stellar-object
(QSO) for optically-selected sources.
There were several motivations for
defining the sample.  First, we were interested in using quasars for
gravitational lensing studies.  A proper determination of lensing
statistics requires the identification of complete quasar samples, as
well as an understanding of any selection effects which might bias
selection of gravitationally lensed quasars. Secondly, the recent
completion of the Large Bright QSO Survey (LBQS: Hewett, Foltz \&
Chaffee 1995) has meant that there is a well-defined sample of
optically-selected QSOs, allowing the determination of global
spectroscopic properties.  The completion of a comparable sample of
radio-selected quasars will allow a detailed phenomenological
comparison of the optical spectra of these two classes of object,
perhaps allowing the determination of differences in underlying
physical conditions.  Finally, quasars are one of the most effective
probes of the universe to high redshift, providing a measure of
evolution as well as the formation of large-scale structure.  Of course
the complete identification of a sample of radio sources can also
provide some surprises, if unexpected objects, such as very high
redshift quasars, are found.

The Parkes Half-Jansky Flat-spectrum Sample we define here contains
323 sources selected in an area of 3.90\Sr\ and is similar to the
earlier compilation by Savage et al.\ (1990).  We have made
significant progress in the optical identification of the sources
which were previously termed ``Empty Fields'', particularly by using
near infrared \kn\ band (2.0--2.3 microns) imaging to detect the
optically faint sources. The extremely red optical to near infrared
colours of these sources imply that most are heavily reddened, viewed
through dust either in the line-of-sight to the quasar or within the
immediate quasar environment (see Webster et al.\ 1995).  In this
paper we present optical identifications for 321 sources (99\%
of the sample), and redshifts for 277 sources (86\%).

The outline of the paper is as follows.  The selection criteria for
the radio sources are described in Section~\ref{sec_sample}.  In
Section~\ref{sec_positions} we explain how the accurate radio positions
were obtained, and present radio images of the resolved sources in the
sample.  Section~\ref{sec_optical} describes the mapping of the
accurate radio positions onto the optical catalogues. We present a
full discussion of the accuracy of this procedure, locate the likely
optical counterparts, classify these images as either stellar or
non-stellar and provide the optical magnitudes.  Where there is no
optical survey image at the location of the radio source, we use
R-band CCD frames and \kn\ near infrared images to determine the source
identification and
morphology.  In Section~\ref{sec_spec} we present spectroscopic
classifications and redshifts of these sources; for those sources
which do not have a published spectrum, we also include the spectra.
All these results are summarised in a master catalogue of the sample in
Section~\ref{sec_cat}. Finally,
Section~\ref{sec_corr} presents a summary of the most important
features of our sample.  An electronic version of our catalogue is
available from the Centre de Donn\'ees astronomiques de Strasbourg
in Section VIII (``Radio Data'') of the catalogue archive.

\section{Selection of the Sample}
\label{sec_sample}

\subsection{Radio Surveys}

Our basic selection criteria are very similar to those used by Savage et
al.\ (1990). We started with the machine-readable version of the Parkes
catalogue (PKSCAT90, Wright \& Otrupcek 1990) and applied the following
criteria:

\begin{enumerate}
\item 2.7\GHz\ ($S_{2.7}$) and 5.0\GHz\ ($S_{5.0}$) fluxes defined
\item $S_{2.7} >  0.5$\Jy\ 
\item spectral index $\alpha_{2.7/5.0} > -0.5$, where $S(\nu) \propto \nu^{\alpha}$
\item Galactic latitude $|b|>20\d$
\item $-45\d<{\rm Declination~(B1950)}<+10\d $
\end{enumerate}

In our search of PKSCAT90, three objects did not have a 5.0\GHz\ flux
defined but satisfied all the other criteria. One of these was later
included from our search of the discovery papers, but the other
objects were not measured at 5.0\GHz\ in the discovery papers,
presumably because they are very bright radio sources associated with
bright optical galaxies: PKS~0131$-$367 (5.6\Jy, 15mag) and PKS~0320$-$374
(98\Jy, 10mag) and were too extended to measure properly. This
search resulted in an initial sample of 325 objects.

We then carefully checked all the radio fluxes in the original
discovery papers of the radio survey, as listed in
Table~\ref{radio_surveys}.  Many objects have more recent (but
unreferenced) flux measurements listed in PKSCAT90, but we replaced
these with the original fluxes in order to quantify the time
difference between the 2.7 and 5.0\GHz\ measurements and thus estimate
the effects of variability. After the original fluxes were adopted, 15
sources no longer satisfied the flux and spectral criteria and so were
excluded. We also found 17 sources whose original fluxes in the
discovery papers satisfied the selection criteria so these were added
to the sample.

In two regions (samples A and F; see Table~\ref{radio_surveys}) our
search of PKSCAT90 produced several sources not listed in the original
papers. These two regions of the original survey were not complete
because the flux limit was not well-defined; subsequent unpublished
observations detected additional sources satisfying our selection
criteria that were included in PKSCAT90. We retained these additional
objects (12 in each region), but flagged them with a minus sign in
front of the reference code (Rf) in the final catalogue
(Table~\ref{tab_master}).

\begin{figure}
\epsfxsize=\one_wide \epsffile{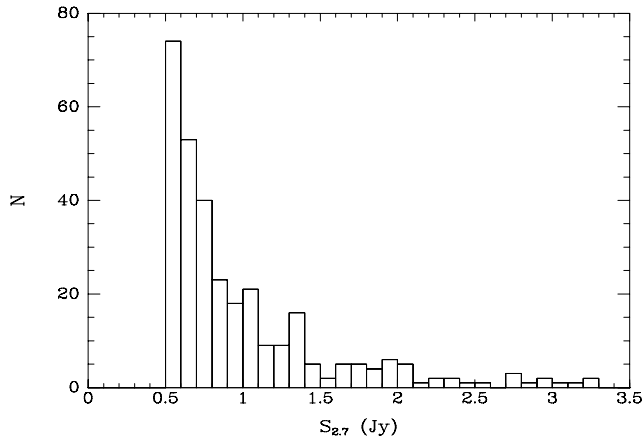}
 \centering
 \caption{Histogram of the 2.7\GHz\ fluxes of the sources in the sample.}
 \label{fig_flux}
\end{figure}

\begin{figure}
\epsfxsize=\one_wide \epsffile{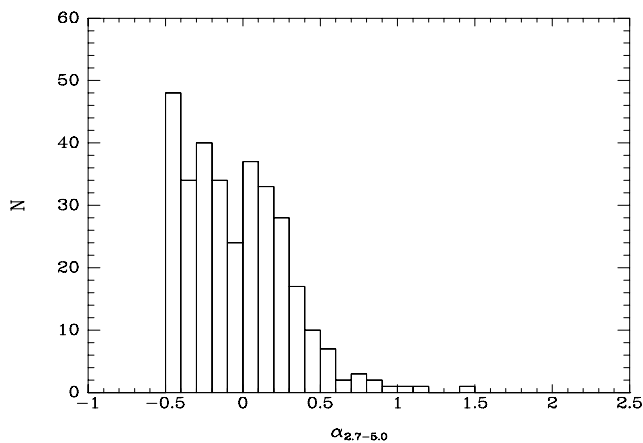}
 \centering
 \caption{Histogram of the (2.7 to 5.0\GHz) radio spectral
indices of the sample sources($S(\nu) \propto \nu^{\alpha}$).}
 \label{fig_alpha}
\end{figure}

\begin{figure*}
\epsfxsize=16.5cm \epsffile{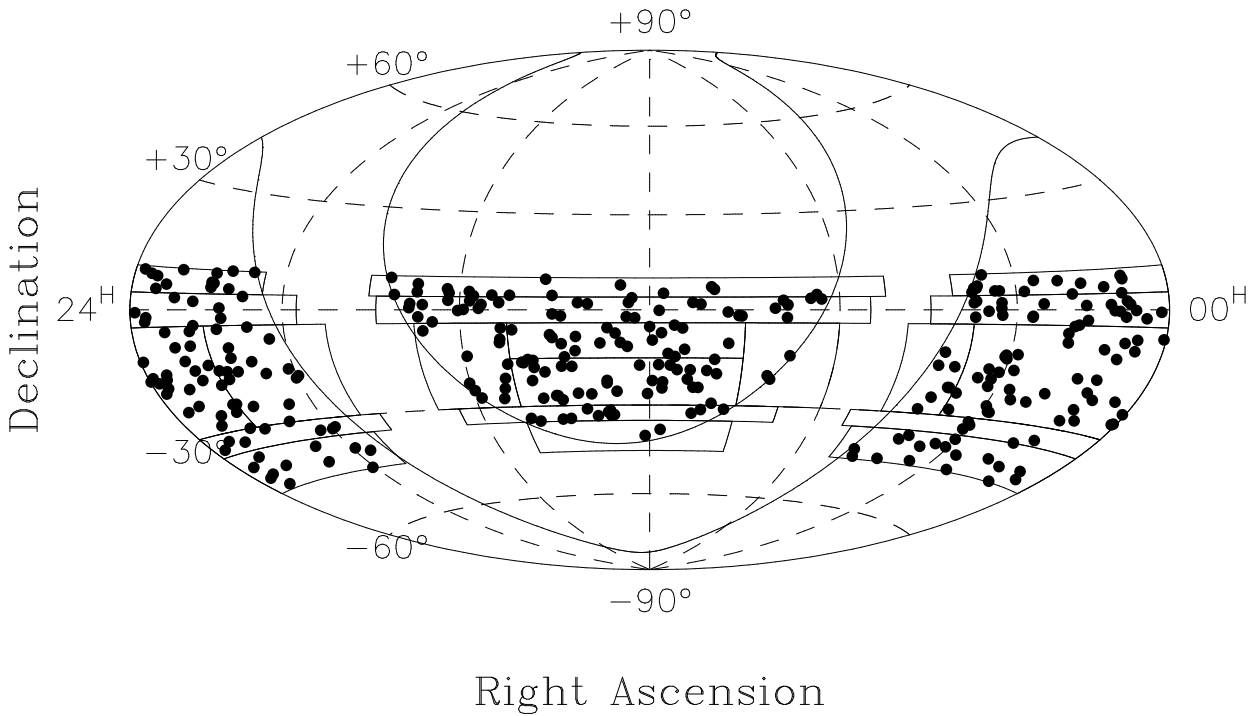}
 \centering
 \caption{Distribution on the sky of The Parkes Half-Jansky Flat-Spectrum Sample
(equal-area projection). The solid lines indicate the the survey regions
and the limits of Galactic latitude ($|b|>20\d$).}
 \label{fig_sky}
\end{figure*}

Finally we removed four planetary nebulae from the sample on the basis
that we are interested in extragalactic sources. This gave a final
sample of 323 sources which are listed in Table~\ref{tab_master} in
Section~\ref{sec_cat}. Our new sample is complete in 6 of the 8
sub-regions listed in Table~\ref{radio_surveys} but in two of the
regions (A and F) the original surveys are incomplete and we have
added additional sources from PKSCAT90. The distributions of the
fluxes and spectral indices are given in Figs. \ref{fig_flux} and
\ref{fig_alpha} respectively and a diagram showing the regions
surveyed and the distribution of our sample across the sky is shown in
Fig.~\ref{fig_sky}.

\begin{table*}
\caption{The Parkes Survey Regions}
\label{radio_surveys}
\noindent
\begin{tabular}{rrrrrlr}
Region &Dec range & RA range & $\Delta T^1$ & Flux Limit & Reference (code)  & $N^2$  \\
       &          &          &    month     & $S_{2.7}$(Jy) \\
\\
A&$+10\d,+04\d$ & 7\h-18\h,20\h30-5\h30 & 2,9  & $\ax0.5^3$ & Shimmins, Bolton \& Wall (1975) (79)   & 46 \\
B&$+04\d,-04\d$ & 7\h20-17\h50,19\h40-6\h & 2-9  & 0.35     & Wall, Shimmins \& Merkelijn (1971)$^4$ (102)  & 63  \\
C&$-04\d,-15\d$ & 10\h-15\h               & 2-14 & 0.25     & Bolton, Savage \& Wright (1979)   (8)  & 21  \\
D&$-15\d,-30\d$ & 10\h-15\h               & 3    & 0.25     & Savage, Wright \& Bolton (1977)   (70)  & 34 \\
E&$-04\d,-30\d$ & 22\h-5\h                & 1-12 & 0.22     & Wall, Wright \& Bolton (1976)     (103)   & 65 \\
F&$-04\d,-30\d$ & 5\h-6\h30,8\h-10\h,15\h-17\h,19\h-22\h 
                                          & 2-9  &$\ax0.6^3$& Bolton, Shimmins \& Wall (1975)   (7)    &39 \\
G&$-30\d,-35\d$ & 9\h-16\h30,18\h30-7\h15 & 1-10 & 0.18     & Shimmins \& Bolton (1974) (78)& 25 \\
H&$-35\d,-45\d$ & 10\h-15\h,19\h-7\h      & 9    & 0.22     & Bolton \& Shimmins (1973) (6) & 30 \\
&&                                                                             &&&total  & 323 \\
\\
\end{tabular}

Notes: 1. $\Delta T$ is the time delay between the 2.7 and 5.0\GHz\
measurements. 2. $N$ is the number of sources each region contributes
to our sample. 3. No completeness analysis was made for regions A and
F so extra objects from PKSCAT90 not in the original papers were
included (12 in each case) and the flux limits are only indicative.
4. The 5\GHz\ fluxes for region B were published separately by Wall
(1972).
\end{table*}

\subsection{Variability}  

Flat-spectrum radio sources are well-known to be variable, which
introduces two biases in our sample. First, our sample was selected
to have a 2.7\GHz\ flux above 0.5\Jy\ {\em at the observation epoch}.
Some of the sample may have been in a particularly bright state; their
average fluxes may be below our limit. Likewise some flat-spectrum
sources with average fluxes above 0.5\Jy\ may have been excluded from
the sample because they were in a particularly faint state when the
sample was defined. 
Secondly, the 5\GHz\ observations of the sample
sources were not obtained simultaneously with the 2.7\GHz\ observations
(see Table~\ref{radio_surveys}). The 5\GHz\ observations were usually
taken after the 2.7\GHz\ observations; the time interval being more than
6 months in $\sim 40$\% of cases; six months is a typical
variability timescale for compact radio sources (Fiedler et al.
1987).  If a source varied between the two observations, its spectral
index could be in error, and the object might be wrongly included in,
or excluded from, the flat-spectrum sample.

Stannard \& Bentley (1977) investigated the variability of 50 Parkes
flat-spectrum radio sources, substantially overlapping our sample. They
compared 2.7\GHz\ fluxes taken two years apart, and found that $\sim
50$\% of sources had varied by 15\% or more.  The number of sources
included in the flux limited sample because they were brighter than
average at the time of observation will exceed the number of sources
missed because they were fainter than average. This is because there
are more sources with mean fluxes just below 0.5\Jy\ that there are with
fluxes just above 0.5\Jy, due to the steepness of the number/flux
relation. Using Stannard \& Bentley's numbers, we estimate that $\sim$
30--40 of our sources have mean fluxes below 0.5\Jy, and that we missed
$\sim$ 20--30 sources with mean fluxes above 0.5\Jy.

Allowing for the time delay between the 2.7\GHz\ and 5\GHz\ measurements,
we can also estimate that $\sim 10$ flat-spectrum sources with $-0.5 <
\alpha < -0.3$ will have been mistakenly classified as steep-spectrum
and excluded from our sample, while another $\sim 10$ with $-0.7 <
\alpha < -0.5$ will have been wrongly included.  This calculation
ignores the dependence of variability on spectral index. Fiedler et
al.\  (1987) showed that most bright compact radio sources with
relatively steep-spectra ($\alpha < -0.2$) vary by only $\sim 5$\% on
timescales of two years. This implies that we will only misclassify
$\sim 5$ objects with $-0.5 < \alpha < -0.3$ as being steep-spectrum.
However, they also find that a small fraction of very flat-spectrum
sources ($\alpha > -0.2$) can vary by 50\% or more on timescales of two
years.  Applying their numbers to our sample, we estimate that $\sim 2$
sources with $\alpha > -0.2$ may have varied by enough to have been
misclassified as steep-spectrum. These numbers may be an overestimate;
Fiedler et al.\ only considered compact sources, whereas several of our
objects, particularly those with steeper spectra, are extended and may
be less variable. We plan to address this uncertainty by remeasuring
the sample making simultaneous flux measurements at both frequencies.

In summary, variability imposes an uncertainty on our 0.5\Jy\
completeness limit at 2.7\GHz: some 30--40 ($\sim$11\%) sources in our
sample have mean fluxes below the limit, and we missed some 20--30
($\sim$8\%) sources with mean fluxes above the limit.  This bias is
inherent to any single-epoch flux-limited sample.

On the other hand we find that $\sim$5--10 objects in our sample
actually have $\alpha <-0.5$ (steep-spectrum) and have been wrongly
included because they varied between the epochs of the 2.7\GHz\ and
5.0\GHz\ measurements, but that another $\sim$5--10 flat-spectrum
objects were missed for the same reason.

\section{Radio Positions}
\label{sec_positions}
 
We had to improve on the poor (10--20 arcsec) accuracy of the original
Parkes radio positions before being able to make optical
identifications of the radio sources by positional coincidence. To
this end we have obtained more accurate radio positions for all
sources in the sample using published data, The VLA Calibrator Manual
(as compiled by Perley \& Taylor, 1996)
and our own Very Large Array (VLA) and Australia
Telescope Compact Array (ATCA) observations. The sources of these
positions and the associated errors are listed in
Table~\ref{tab_Rradio}. The source positions are listed in
Table~\ref{tab_master}; note that we use the original naming scheme
for the sources based on B1950 coordinates but we include the J2000
coordinates for all the sources in Table~\ref{tab_master} for
reference.

\begin{table}
\centering
\caption{Sources of accurate radio positions}
\label{tab_Rradio}
\begin{tabular}{lr}
 Reference (code)                & uncertainty   \\
                                 & (arcsec) \\
\\
 Jauncey et al.\ (1989)   (39)   &  0.15      \\
 Johnston et al.\ (1995)  (40)   &  0.01      \\
 Lister et al.\ (1994)    (43)   &  $\ax$0.3  \\
 Ma et al.\ (1990)        (45)   &  0.01      \\
 Morabito et al.\ (1982)  (50)   &  0.6       \\
 Patnaik (1996)           (55)   &  $\ax$0.02 \\
 Perley (1982)            (56)   &  0.15      \\
 Perley \& Taylor (1996)  (57)   &  $\ax$0.15 \\
 Preston et al.\ (1985)   (60)   &  0.6       \\
 Ulvestad et al.\ (1981)  (97)   &  0.40      \\
 This paper: ATCA        (120)   &  0.0-0.3   \\
 This paper: VLA         (121)   &  0.2-0.5   \\
\end{tabular}
\end{table}

\subsection{VLA Observations and Data Reduction}

On 1986 October 1 and 4 we observed the majority of the sources that
lacked accurate published positions with the VLA. The observations were
made at 4.86\GHz\ with the VLA in its ``CnB'' configuration to yield
nearly circular synthesised beams with approximately 6 arcsec FWHM
resolution.  Each programme source was covered with a single ``snapshot''
scan of about 3 minutes duration, and each group of snapshots was
preceded and followed by scans on a phase calibrator whose rms absolute
position uncertainty is not more than 0.1 arcsec in each coordinate.
The phase calibrator flux densities were bootstrapped to the Baars et
al.\ (1977) scale via observations of 3C 48 and 3C 286.

The (u,v) data recorded from both circular polarizations in two 50 \MHz\ 
bands centered on 4.835 and 4.885\GHz\ were edited, calibrated, and
mapped with AIPS.  The images were cleaned, and the clean components
were used to self-calibrate the antenna phases, yielding images with
dynamic ranges typically $> 200:1$.  Nearly every programme source
contains a dominant compact component that should coincide in position
with any possible optical identification.  The positions of these
compact components were determined by Gaussian fitting on the images.
The formal fitting residuals are $< 0.1$ arcsec because the synthesised
beam is small and the signal-to-noise ratios are high.  Thus the radio
position uncertainties are dominated by atmospheric phase drifts and
gradients not removed by the calibration.  They range from about 0.2
arcsec at Dec $+10\d$ to about 0.5 arcsec at Dec $-45\d$.

\subsection{ATCA Observations and Data Reduction}


Several remaining sources in the sample were observed with the ATCA
during 1993 March and November using all 6 antennas with a maximum
baseline of 6~km. Observations were made at 4.80 and 8.64\GHz\ in ``cuts''
mode with orthogonal linear polarisations at a bandwidth of
128\MHz. The synthesised beam at 4.80\GHz\ has a constant East-West
resolution of 2 arcsec FWHM and a North-South resolution varying
from 3 arcsec (at Dec $-45\d$) to 8 arcsec (Dec $-21\d$).
``Cuts'' mode involves observing each object for a period of one
minute on at least 6 occasions spread evenly over 12 hours. In this
way, it is possible to obtain imaging data on approximately 40 sources
within a 12 hour observation. Secondary calibrators with accurate,
milliarcsec positions close to the programme sources were observed
at least once every 2 hours. The flux density scale was determined
from observations of the primary calibrator at the ATCA, PKS~1934$-$638.

The data were edited and calibrated within AIPS and images made using
the Caltech Difmap software (Shepherd, Pearson \& Taylor, 1995). The
final self-calibrated images have typical dynamic ranges in excess of
400:1 for strong and relatively compact sources, decreasing to
approximately 100:1 for objects with weak or extended emission.
Source positions were calculated by fitting a Gaussian to the peak in
the brightness distribution of a cleaned (but not self-calibrated)
8.64\GHz\ image.  The uncertainty in source positions measured from these
ATCA images comprises a component due to thermal noise, which scales
inversely with S/N ($\sim$beamwidth/(S/N) ) and a component due to
systematic effects arising from the phase-referencing. The latter term
dominates for strong sources and scales linearly with angular distance
between the source and the phase-reference used to calibrate its
position. The error is approximately 0.1 arcsec for an angular
separation of 5\d\ (Reynolds et al.\ 1995).

\subsection{The Radio Positions and Morphology}

The new radio positions are presented in Table~\ref{tab_master}. As
shown in Table~\ref{tab_Rradio} these are all accurate to 0.6
arcsec or better for unresolved sources. Any radio sources we know
to be resolved are noted in the comments column of
Table~\ref{tab_master} and we present radio images of these sources in
Fig.~A1.  We indicate five different categories of
resolved source in the Table using the terminology of Downes et al.
(1986):

\begin{enumerate}

\item ``P'' signifies partially resolved sources: the position is 
well-defined by a peak.

\item ``Do'' indicates double sources with no central component or dominant
peak. There is no clear maximum, so the centroid of the image was used to
define the position.

\item ``Do+CC'' indicates a double-lobed source with a central component
or peak that gives a well-defined position.

\item ``H'' indicates a diffuse halo around a central source which
gives a well-defined position.

\item ``HT'' indicates a complex head-tail structure with no
well-defined position.
\end{enumerate}

\subsection{Notes on Individual Radio Positions}
\label{sec_notes}

In this section we describe any sources with extended structure
making the position difficult to define. We also note any sources
for which our final accurate positions differ by more than 24 arcsec
from the original Parkes catalogue positions.

\begin{enumerate}

\item PKS~0114+074: there are 3 components to the VLA radio image in
Fig.~A1. We have adopted the centroid of the stronger
double source to the South, although the Northern source also has an
optical counterpart. The PKSCAT90 position corresponds to the Northern
source; our position is therefore some 30 arcsec different. Our
spectroscopic observations show that the Northern source (at
01:14:49.51 $+$07:26:30.0 B1950) is a broad-lined quasar at $z=0.858$
consistent with previous publications.  The correct identification (at
01:14:50.48 $+$07:26:00.3 B1950) is a narrow-line galaxy at $z=0.342$.

\item PKS~0130$-$447: this position is some 30 arcsec from the 
original value.

\item PKS~0349$-$278: the VLA image in Fig.~A1 is confused
with a compact source some 2.5 arcmin from the PKSCAT90 position
and a marginal detection at the PKSCAT90 position. We made an
independent check of the radio centroid position for this source by
measuring it on the 4.85\GHz\ survey images made with the NRAO 140-foot
telescope (Condon, Broderick \& Seielstad, 1991).  A Gaussian fit gave
a position of 03:49:31.5 $-$27:53:41 (B1950), consistent with the
original position (and coincident with an optical galaxy) but not with
the stronger VLA source at 03:49:41.17 $-$27:52:07.0 (B1950).
Furthermore, the fit is clearly extended (source size 280 arcsec by 109
arcsec with position angle 50\d\ after the beam has been deconvolved).
The 4.85\GHz\ flux of PKS~0349$-$278 is just over 2\Jy, but the strong
source in the VLA image is only 0.3\Jy.  The VLA has resolved out most
of the flux, leaving only two components plus some residuals visible in
the contour plot.  The strong VLA component is probably only a hotspot
in the northeastern lobe of the radio source. We adopt the fainter VLA
position (03:49:31.81 $-$27:53:31.5 B1950) which is consistent with the
single-dish positions.

\item PKS~0406$-$311: the VLA image in Fig.~A1 shows a
complex head-tail source with no clear centre. The Northern limit of
the source is close to a bright galaxy. We tentatively claim this as
the identification, although the separation is 7.25 arcsec from
the poorly defined ``head'' of the radio source and about 35
arcsec from the original position.

\item PKS~0511$-$220: we find a very large difference between our
position for this source (05:11:41.81 $-$22:02:41.2, B1950) and that
quoted in Hewitt \& Burbidge (1993) (05:11:49.94 $-$22:02:44.8). We
attribute this difference to a typographical error made with respect
to the position (05:11:41.94 $-$22:02:44.8) given by Condon, Hicks \&
Jauncey (1977). We are concerned that any published redshifts of this
object may correspond to an object near the wrong position so we do
not quote a redshift for this source pending our own observations.

\item PKS~1008$-$017: (see Fig.~A1) our new position is
about 40 arcsec from the original value.

\item PKS~1118$-$056: this is 60 arcsec away from the original survey 
position; we suspect a typographical error in the discovery paper
(Bolton et al.\ 1979).

\item PKS~2335$-$181: in the case of this double source (see
Fig.~A1) with no central component the centroid of the
image was not used to define the position; the North-East component
was adopted instead. This was chosen because of the very good
positional correspondence with a quasar at redshift z=1.45 and also
the fact that no optical counterpart for the South-West component was
detected in the Hubble Space Telescope Snapshot Survey (Maoz et al.\
1993).

\end{enumerate}

\section{Optical Identifications}
\label{sec_optical}

\subsection{Matching to Sky Survey Positions}

A major advance that we present in this paper is the matching of our
accurate radio source positions to the accurate optical data now
available in large digitised sky catalogues based on the U.K. Schmidt
Telescope (UKST) and Palomar sky surveys. A factor contributing to
our successful identifications is the greatly improved agreement
between the radio and optical reference frames in the South (e.g.
Johnston et al.\ 1995).

Our major source of optical data is the COSMOS/UKST Southern Sky
Catalogue. This lists image parameters derived from automated
measurements of the ESO/SERC Southern Sky Survey plates, taken on
IIIa-J emulsion with the GG395 filter to give the photographic blue
passband \Bj\ (3950--5400\AA). The catalogue is described further by
Yentis et al.\ (1992).  There are systematic errors in the astrometry
of the COSMOS catalogue:  we made a first-order correction as described
by Drinkwater, Barnes \& Ellison (1995) by using the PPM star catalogue
(R\"oser, Bastian \& Kuzmin 1994) to calculate a mean shift in the
positions for each Schmidt field used.

For sources North of +3 degrees we used data from the Automated Plate
Measuring facility (APM; see Irwin, Maddox \& McMahon 1994) at
Cambridge based on blue (unfiltered 103a-O emulsion; 3550--4650\AA)
and red (red plexiglass 2444 filter plus 103a-E emulsion;
6250--6750\AA) plates from the first Palomar Observatory Sky Survey
(POSS~I).

The sky catalogues were used to generate finding charts for all the
sources which we present in Appendix~\ref{sec_charts}. These charts
are a good approximation to the photographic data, but we stress that
there can be problems with image merging in crowded fields: close
objects (e.g. two stars) can be misclassified as a ``merged'' object
or galaxy. The ``Field'' code at the bottom of each
chart indicates the UKST field number (or the plate number for POSS~I)
with a prefix describing the type of plate. The prefix ``J'' indicates
UKST \Bj\ plates measured by COSMOS. For APM data ``j'' indicates UKST
\Bj\ plates, ``O'' blue POSS~I plates and ``E'' red POSS~I plates.

\begin{figure}
\epsfxsize=\one_wide \epsffile{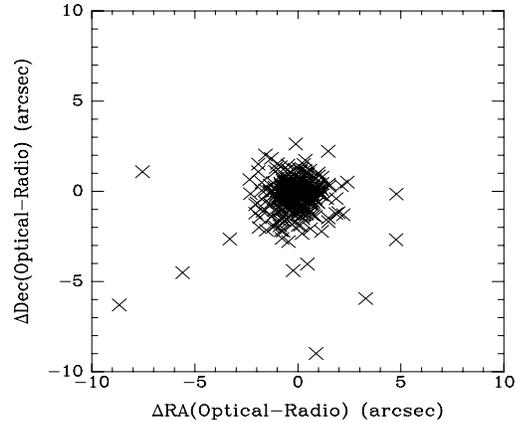}
 \centering
 \caption{Distribution of the Position Offsets between each radio
source position and the nearest detected image in the sky catalogues.}
 \label{fig_scatter}
\end{figure}

The procedure to find the optical counterpart to each radio source
started with the selection of the nearest optical image in the
catalogues to each radio position. The relative positions of these
nearest-neighbours are shown in Fig.~\ref{fig_scatter}. There is a
clear concentration at small separations (less than 3 arcsec), but
we note that in some cases the nearest-neighbours are at larger
separations (greater than 5 arcsec).  We made a preliminary
estimate of the spread in the position offsets by fitting Gaussians to
the distributions in RA and Dec; the rms scatter was found to be
about 0.9 arcsec in each direction. A preliminary cutoff separation of
4 arcsec (about 4$\sigma$) was then imposed.

\begin{figure*}
\epsfxsize=15cm \epsffile{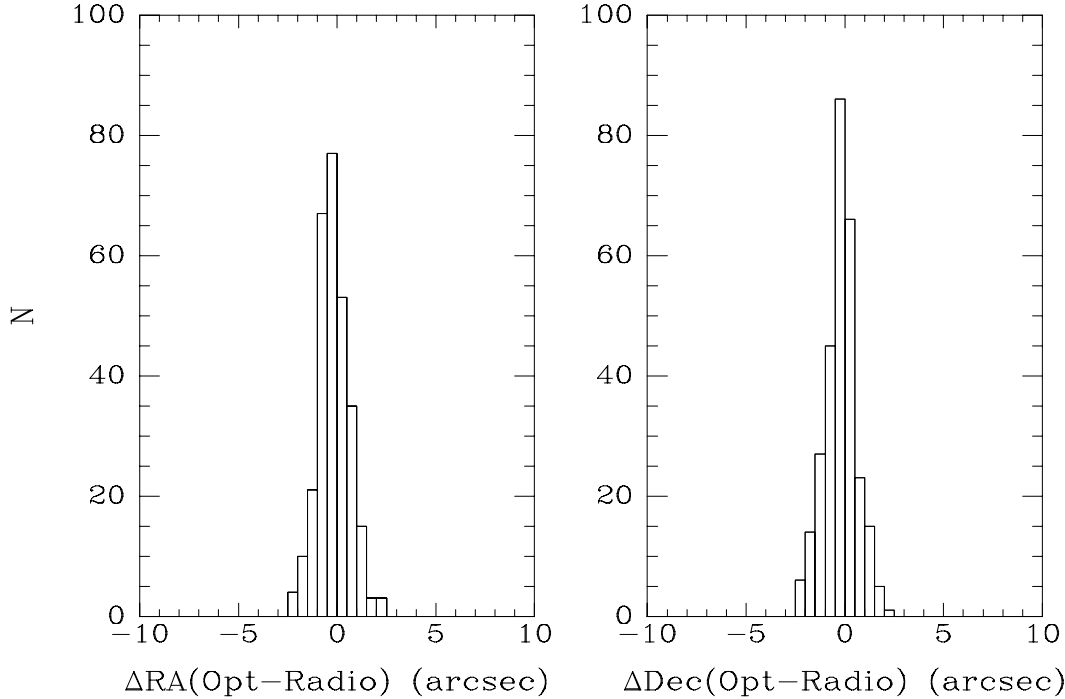}
 \centering
 \caption{Histograms of the Position Offsets between the Radio and the
 Optical sources.}
 \label{fig_histo}
\end{figure*}

We removed the outliers more distant than 4 arcsec and then
recalculated the distributions of position offsets: these are shown
plotted in Fig.~\ref{fig_histo} as histograms of the offsets between
the two positions in RA and Dec. We estimated the statistical range of
this distribution by measuring the Gaussian dispersions in RA and Dec.
These results are given in Table~\ref{tab_offsets}. This shows that
the core of the distribution has dispersions of only about 0.8
arcsec in each direction.  (The same table also shows the final
results with fainter objects matched on CCD frames included.)  The
mean differences are small (about 0.2 arcsec) but significant (4 sigma
formally) in both RA and Dec: these indicate that some residual
systematic effects remain, mostly likely due to remaining second-order
errors in the COSMOS astrometry. The important point is that the small
dispersion in both measurements allows us to place very strong limits
on the identification of our sources.

We adopted a maximum difference of $\pm2.5$ arcsec in RA and $\pm2.5$
arcsec in Dec between the radio source position and nearest optical
image, corresponding to a $3\sigma$ confidence level in each
coordinate. We did not remove the small systematic mean offsets before
applying these limits.  The maximum total separation among the
objects satisfying these criteria was 2.7 arcsec.  In all cases where
the matching criteria were satisfied the image parameters from the
automated catalogues are given in Table~\ref{tab_master}: the
optical$-$radio position offsets in arcsec, the morphological
classification and the catalogue \bj\ magnitude.

\begin{table}
\caption{Mean Optical--Radio Position Offsets}
\label{tab_offsets}
\begin{tabular}{lrrrrr}
sample             &  N  &  $\overline{\Delta\ra}$  &  $\sigma_{RA}$ 
                         &  $\overline{\Delta\dec}$ &  $\sigma_{Dec}$ \\
		   &     & arcsec & arcsec          & arcsec  & arcsec \\
\\
sky survey matches & 290 &  -0.17 &     0.82        &  -0.21   & 0.81  \\
all matches        & 320 &  -0.16 &     0.83        &  -0.18   & 0.82  \\ 
\\
\end{tabular}

Note: each offset is measured in the sense optical$-$radio and
PKS~0406$-$311 is not included.

\end{table}

The morphological classifications are based on how extended the
optical images are and define the images as galaxies (g), stellar (s),
or too-faint-to-classify (f). In the case of the APM data there is a
further category of merged images (m) where 2 or more images are too
close to separate.  We intentionally do not include in
Table~\ref{tab_master} the object classifications from PKSCAT90
because there is evidence that the distinction between ``galaxies'' and
``quasars'' was not applied uniformly over the whole survey (see
Drinkwater \& Schmidt 1996).

\begin{figure}
\epsfxsize=\one_wide \epsffile{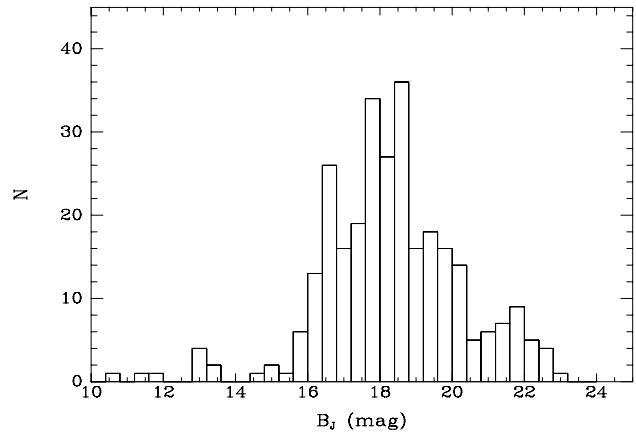}
 \centering
 \caption{Histogram of the optical \bj\ magnitudes of all sources identified
on the photographic sky surveys.}
 \label{fig_mags}
\end{figure}

The calibration accuracy of the \bj\ photographic magnitudes from the
COSMOS catalogue is quoted as being about $\pm0.5$ magnitudes (H.
MacGillivray, private communication). We have found that some fields
lack any calibration data and some seem to be incorrect by more than
one magnitude, so the catalogue magnitudes should be treated with
caution.  We specifically checked the calibration of any fields in
which the COSMOS magnitude differed by more than 2 mag from a value
published in the literature by comparison with data from adjacent
COSMOS fields and corrected any large errors.  A histogram of the
magnitudes is given in Fig.~\ref{fig_mags}.

There is a further problem of objects where 2 or 3 close optical
images have been merged into a single catalogue object whose centroid
position is still within our $\pm2.5$ arcsec matching criteria.
These are easy to find because the resulting ``merged'' image is very
extended and thus mis-classified as a galaxy. This is a problem
inherent in automated catalogues for which reason the classifications
should always be checked. We visually inspected all the objects
classified as ``galaxies'' to check for merging. A total of 10 such
objects were found in the matched list; they are noted in
Table~\ref{tab_master} as ``(merge)''. We derived corrected image
parameters for these merged objects by analysing images from the
Digitised Sky Survey or CCD images (at other wavelengths, see next
section). If the image data was obtained from CCD data, no
\bj\ magnitude is given for the object in the table.

%

One additional object was included in the matched list although its
position difference was greater than the $\pm2.5$ arcsec
limits. This was PKS~0406$-$311 which we identified with a galaxy 7
arcsec from the nominal radio position because the head-tail radio
structure did not give an accurate position but is very indicative of
this type of galaxy (see above). With this galaxy and the merged
objects included, a total of 291 sources from our sample of 323 have
confirmed matches to objects in the optical catalogues.  
This leaves a total of 32
sources with no matching image in the optical catalogues. We
undertook the identification of these sources using CCD imaging at
other wavelengths as described in the next section.

\subsection{Identifications at Other Wavelengths}

This section describes the methods we used to find optical
counterparts for the 32 sources not matched to images listed in the
optical catalogues.  We first inspected all the fields visually on the
optical sky survey plates. Most of the sources (25) were found to be
genuine ``empty fields'' in the sense that no optical counterpart was
visible on the survey plates. In the remaining cases (7) however a
counterpart was clearly visible on the plate, but it was too faint to
be included in the automated catalogue or it had been merged with a
neighbouring object.

We note that six of these unmatched sources were assigned optical
identifications in PKSCAT90; our new accurate positions show that these
need to be revised. In three cases (PKS~1349$-$145, PKS~1450$-$338,
PKS~2127$-$096) there is a faint matching object but it is merged with a
brighter image and in the other cases there is no optical counterpart
at all at the correct position (PKS~0005$-$262, PKS~1601$-$222,
PKS~2056$-$369).

To identify the sources unmatched on the sky surveys we turned to longer
wavelengths, using optical \r-band, \i-band, and 
infrared \kn-band (2.0--2.3
microns) imaging on the
3.9m Anglo-Australian Telescope
(AAT) and the 
Australian National University (ANU) 2.3m Telescope. These data were
analysed using the {\footnotesize IRAF}\footnote{IRAF is distributed
by the National Optical Astronomy Observatories, which are operated by
the Association of Universities for Research in Astronomy, Inc. (AURA)
under cooperative agreement with the National Science Foundation.}
analysis software.  The observations resulted in identifications of 30
of the remaining sources including the merged objects, one of which
was separated using a \b-band image. These sources are listed in
Table~\ref{tab_master} in the same way as the sources identified from
the digitised survey data, except that no \Bj\ magnitude is given and
the position offsets are estimated from the CCD frames. The source of
the identifications is indicated in the comment column as ``(R)'' or
``(K)''. We will present a full analysis of the \r- and \kn-band
data in later papers. 

The 2 sources we did not identify include PKS~1213$-$172 which lies
too close to a bright star to be identified in our data but Stickel et
al.\ (1994) report having identified it with a ``$m=21.4$ mag resolved
galaxy''. The remaining source, PKS~0320$+$015 was not detected in a
\kn\ image (approximate limit of \kn=18) but we anticipate identifying
it when a deeper exposure is available.

We note that PKS~2149+056 which we detected in our \kn\ image
was previously detected and identified as a
quasar with a measured redshift by Stickel \& K\"uhr (1993).

\subsection{Reliability of Identifications}

For the majority of the matched sources for which spectroscopic redshifts
have been measured we are confident of having made the correct optical
identification. For the remaining sources for which we have not yet obtained
redshifts, the identifications must be made on positional coincidence alone.
A very detailed analysis of the statistics of source identifications was 
made by Sutherland \& Saunders (1992) in the context of matching IRAS sources
with poor positions to the optical sky survey data. Our problem is much simpler
because both our source (radio) and survey (optical) positions are accurate.
Furthermore, we do not wish to include the image magnitudes in the analysis
because we do not know the true distribution of optical magnitudes---a large
fraction of the sources without spectroscopic confirmation are at the faint
limit of the magnitude distribution.

We made an estimate of the number of ``identifications'' in our sample
that might just be coincidences by calculating the mean surface density
of images in the sky survey catalogues at the plate limit and finding
how many of these would lie within the match criteria. For the 46
fields without spectroscopic confirmation we would expect 1 random
matches within a radius of 3 arcsec.  In fact most objects lie
within 2 arcsec: at this separation we would only get 0.4 random
matches. It is therefore possible that one of the identifications we
claim without spectroscopic confirmation is wrong: ideally only the
sources with spectroscopic identifications should be used for analysis
purposes.

\section{Spectroscopic Identifications}
\label{sec_spec}

\subsection{Previous Results}

Earlier versions of the flat-spectrum sample have been the subject of
extensive campaigns of spectroscopic follow-up observations. Some two
thirds of the sample were identified in the summary made by Savage
et al.\ (1990) and we have drawn on this work for the current sample.

We carried out a very detailed literature review to find published
redshifts for as much of the sample as possible. We based our search
on the quasar catalogue compiled by Hewitt \& Burbidge (1993) with
additional material from the V\'eron-Cetty \& V\'eron (1993) quasar
catalogue, the Center for Astrophysics Redshift Catalog (Version of
May 28, 1994; see Huchra et al.\ 1992), the NASA/IPAC Extragalactic
Database (NED, Helou et al.\ 1991), and the Lyon-Meudon Extragalactic
Database (LEDA).  There are occasional errors in some of these large
compilations, so for every redshift found in the catalogues we checked
the reference cited and only accepted values for which we found a
measured redshift in the original reference. We present these
redshifts in Table~\ref{tab_master} along with a code that specifies
the source of the measurement. For some objects the source reference
indicated that the redshift was uncertain (e.g. due to a single line
or a lower limit derived from the redshift of an absorption system);
in these cases the reference code is prefaced by a ``$-$'' sign. For
some additional objects we found no published original reference (some
were given by private communications): these are assigned a reference
code of zero and we have not listed the redshift in our table.

After our critical search of the literature we accepted published
redshifts for 206 sources in our sample of 323 sources. At the same
time we searched for published spectra of any sources in our sample;
references to these are also given in Table~\ref{tab_master}. Again,
we only include those spectra we have checked in the original
references.

\subsection{New Measurements}

As a result of the identifications presented in this paper we started
a campaign of new spectroscopic identifications. This has resulted in
114 new spectra and 90 new redshift measurements which we present
here.  The journal of observations and the new redshifts are given in
Table~\ref{spectral_id} and we present the spectra in
Appendix~\ref{sec_spectra}. Notes on some individual spectra are given
in Section~\ref{sec_snotes} below.  Note that three sources are
presented (``EXTRAS'' in Table~\ref{spectral_id}) that are not in our
final sample.  These were part of an earlier version of the sample and
are included here to provide a published reference to their
redshifts. Details of our observations are as follows.

\begin{table*}
\caption{New Spectral Identifications}
\label{spectral_id}
\begin{tabular}{lrrrllrrrl}
name      & tel & date      & $z_{em}$ & comment &name   & tel & date  & $z_{em}$ & comment \\
\\
PKS~0036$-$216&AAT&1995 Sep 22&none    &     & PKS~1143$-$245&ANU&1995 May 25&1.940   &          \\
PKS~0048$-$097&AAT&1994 Dec 02&none    &     & PKS~1144$-$379&AAT&1996 Apr 21&1.047   &(87)      \\
PKS~0104$-$408&AAT&1984 Jun 30&none    &(105)& PKS~1156$-$094&AAT&1996 Apr 20&none    &          \\
PKS~0114$+$074&AAT&1995 Sep 22&0.343   &note & PKS~1228$-$113&AAT&1996 Apr 21&3.528   &          \\
PKS~0118$-$272&AAT&1994 Dec 04&$>$0.556&     & PKS~1237$-$101&ANU&1995 May 25&0.751   &          \\
PKS~0131$-$001&AAT&1994 Dec 03&0.879   &     & PKS~1250$-$330&AAT&1996 Apr 20&none    &          \\
PKS~0138$-$097&ANU&1995 Sep 28&none    &(90) & PKS~1256$-$229&AAT&1995 Mar 05&1.365   &          \\
PKS~0153$-$410&AAT&1994 Dec 04&0.226   &     & PKS~1258$-$321&AAT&1988 May 10&0.017   &(18)      \\
PKS~0213$-$026&AAT&1994 Dec 04&1.178   &     & PKS~1317$+$019&AAT&1996 Apr 21&1.232   &          \\
PKS~0216$+$011&AAT&1994 Dec 03&1.61    &     & PKS~1318$-$263&AAT&1995 Mar 05&2.027   &          \\
PKS~0220$-$349&AAT&1994 Dec 04&1.49    &     & PKS~1333$-$082&AAT&1988 May 10&0.023   &(26)      \\
PKS~0221$+$067&AAT&1986 Aug 09&0.510   &     & PKS~1336$-$260&AAT&1995 Mar 05&1.51    &note      \\
PKS~0229$-$398&AAT&1994 Dec 04&1.646?  &     & PKS~1340$-$175&AAT&1996 Apr 20&1.50?   &1 line    \\
PKS~0256$+$075&AAT&1994 Dec 03&0.895   &     & PKS~1354$-$174&AAT&1995 Mar 06&3.137   &          \\
PKS~0301$-$243&AAT&1995 Sep 22&none    &     & PKS~1359$-$281&AAT&1984 May 01&0.803   &          \\
PKS~0327$-$241&AAT&1994 Dec 04&0.888   &     & PKS~1404$-$267&AAT&1988 May 10&0.022   &(21)      \\
PKS~0332$-$403&ANU&1995 Sep 27&none    &     & PKS~1406$-$267&AAT&1996 Apr 20&2.43    &          \\
PKS~0336$-$017&AAT&1987 Sep 17&3.202   &     & PKS~1430$-$155&AAT&1996 Apr 21&1.573   &          \\
PKS~0346$-$163&ANU&1995 Sep 28&none    &     & PKS~1435$-$218&ANU&1996 Feb 25&1.187   &          \\
PKS~0346$-$279&AAT&1986 Aug 09&0.987   &     & PKS~1445$-$161&AAT&1984 May 01&2.417   &          \\
PKS~0357$-$264&AAT&1995 Sep 22&1.47?   &     & PKS~1450$-$338&AAT&1996 Apr 20&0.368   &          \\
PKS~0400$-$319&AAT&1994 Dec 03&1.288   &     & PKS~1456$+$044&AAT&1988 May 11&0.394   &          \\
PKS~0405$-$331&AAT&1987 Sep 17&2.562   &     & PKS~1511$-$210&AAT&1994 Apr 30&1.179   &          \\
PKS~0406$-$311&ANU&1995 Sep 27&0.0565  &     & PKS~1518$+$045&ANU&1995 May 25&0.052   &          \\
PKS~0422$+$004&ANU&1995 Sep 28&none    &     & PKS~1519$-$273&ANU&1996 Apr 11&none    &          \\
PKS~0423$+$051&AAT&1994 Dec 02&1.333   &     & PKS~1535$+$004&AAT&1996 Apr 21&3.497   &          \\
PKS~0454$+$066&ANU&1995 Sep 28&0.4050  &     & PKS~1615$+$029&AAT&1996 Apr 21&1.341   &(110)     \\
PKS~0456$+$060&AAT&1995 Mar 05&none    &     & PKS~1616$+$063&AAT&1996 Apr 21&2.088   &(3)       \\
PKS~0459$+$060&AAT&1994 Dec 03&1.106   &     & PKS~1635$-$035&AAT&1988 May 11&2.856?  &          \\
PKS~0502$+$049&AAT&1995 Mar 05&0.954   &     & PKS~1648$+$015&AAT&1996 Apr 20&none    &note      \\
PKS~0508$-$220&ANU&1995 Nov 29&0.1715  &     & PKS~1654$-$020&AAT&1996 Apr 20&2.00    &          \\
PKS~0532$-$378&AAT&1995 Mar 05&1.668   &     & PKS~1706$+$006&AAT&1994 Sep 09&0.449   &          \\
PKS~0829$+$046&AAT&1994 Dec 02&none    &     & PKS~1933$-$400&ANU&1995 May 25&0.965   &          \\
PKS~0837$+$035&AAT&1995 Mar 05&1.57    &     & PKS~1958$-$179&AAT&1996 Apr 21&0.652   &(10)      \\
PKS~0859$-$140&AAT&1996 Apr 21&1.337   &(84) & PKS~2004$-$447&AAT&1984 May 02&0.240   &          \\ 
PKS~0907$-$023&AAT&1995 Mar 05&0.957   &(110)& PKS~2021$-$330&AAT&1996 Apr 21&1.471   &(98) note \\
PKS~0912$+$029&AAT&1988 May 11&0.427   &     & PKS~2022$-$077&AAT&1988 May 10&1.388   &          \\
PKS~0922$+$005&AAT&1995 Mar 05&1.717   &     & PKS~2056$-$369&AAT&1995 Jul 06&none    &          \\
PKS~1008$-$017&ANU&1996 Apr 10&0.887   &note & PKS~2058$-$135&AAT&1988 May 10&0.0291  &(21)      \\
PKS~1016$-$311&AAT&1988 May 10&0.794   &     & PKS~2058$-$297&AAT&1984 May 02&1.492   &          \\
PKS~1020$-$103&ANU&1996 Apr 26&0.1966  &(112)& PKS~2059$+$034&AAT&1996 Apr 21&1.012   &(110)     \\
PKS~1021$-$006&ANU&1996 Apr 26&2.549   &(110)& PKS~2120$+$099&AAT&1987 Sep 17&0.932   &          \\
PKS~1036$-$154&AAT&1995 Mar 05&0.525   &     & PKS~2127$-$096&AAT&1995 Jul 06&$>$0.780&$>$0.733  \\
PKS~1038$+$064&ANU&1996 Apr 26&1.264   &(84) & PKS~2128$-$123&AAT&1996 Apr 21&0.499   &(95)      \\
PKS~1048$-$313&AAT&1995 May 31&1.429   &     & PKS~2131$-$021&ANU&1995 Jun 01&1.285   &note      \\
PKS~1055$-$243&AAT&1995 Mar 05&1.086   &     & PKS~2143$-$156&ANU&1995 May 25&0.698   &          \\
PKS~1102$-$242&AAT&1984 May 01&1.666   &     & PKS~2145$-$176&AAT&1987 Sep 17&2.130   &          \\
PKS~1106$+$023&AAT&1988 May 10&0.157   &     & PKS~2215$+$020&AAT&1986 Aug 10&3.572   &          \\
PKS~1107$-$187&AAT&1995 Mar 05&0.497   &     & PKS~2229$-$172&AAT&1995 Jul 06&1.780   &          \\
PKS~1110$-$217&AAT&1996 Apr 20&none    &     & PKS~2233$-$148&AAT&1995 Jul 06&$>$0.609&          \\
PKS~1115$-$122&AAT&1988 May 10&1.739   &     & PKS~2252$-$090&AAT&1996 Jul 19&0.6064  &          \\
PKS~1118$-$056&AAT&1988 May 11&1.297?  &     & PKS~2254$-$367&AAT&1988 May 11&0.0055  &(21)      \\
PKS~1124$-$186&ANU&1996 Apr 26&1.048   &note & PKS~2312$-$319&ANU&1995 Sep 28&1.323   &$>$1.0453 \\
PKS~1127$-$145&AAT&1996 Apr 21&1.187   &(107)& PKS~2329$-$415&AAT&1987 Sep 17&0.671   &          \\
PKS~1128$-$047&AAT&1984 May 01&0.266   &     & PKS~2335$-$181&AAT&1987 Sep 17&1.450   &          \\
PKS~1133$-$172&AAT&1994 Apr 30&1.024   &     & \multicolumn{4}{c}{EXTRAS}                     \\
PKS~1136$-$135&ANU&1996 Apr 26&0.5566  &(107)& PKS~0114$+$074b&AAT&1995 Sep 22&0.858  &note      \\
PKS~1142$+$052&AAT&1986 Apr 11&1.342   &(105)& PKS~0215$+$015&AAT&1994 Dec 02&1.718   &note      \\
PKS~1142$-$225&AAT&1996 Apr 21&1.141   &     & PKS~1557$+$032&AAT&1995 Mar 5 &3.88    &note      \\ 
 & \\
\end{tabular}
 \\ 

Notes: 1. Specific notes on individual spectra are given in
Section~\ref{sec_snotes}.  2. Redshifts of any absorption systems
identified in the spectra are prefaced by ``$>$'' as these give a
lower limit to the source redshift. 3. Reference numbers in
parentheses refer to previous published redshift estimates.  4. The
final three sources observed are not in our sample but are included
here in order to provide a published reference to their redshifts.

\end{table*}

Our identification of most of the ``Empty Field'' sources in our \kn\
and \r\ band imaging enabled us to attempt spectroscopic
identifications of these very faint sources. We made these observations
using the AAT equipped with the RGO
spectrograph (grating 250B: a resolution of 5\AA\ in the blue) and the
faint object red spectrograph (FORS: a resolution of 20\AA\ in the
red).  We also used the AAT to observe a number of brighter objects
with unconfirmed redshifts.

We made an extensive search of the AAT archive for observations of
sources in our sample with no published redshifts: this provided
27 measurements.

A number of the brighter objects were observed with the ANU 2.3m
Telescope using the double beam spectrograph (with a resolution of
8\AA\ in both the blue and red arms).

All these spectra were analysed with the IRAF package and any new
redshifts we obtained are included in Table~\ref{tab_master} with the
reference code ``121''. 

\begin{figure}
\epsfxsize=\one_wide \epsffile{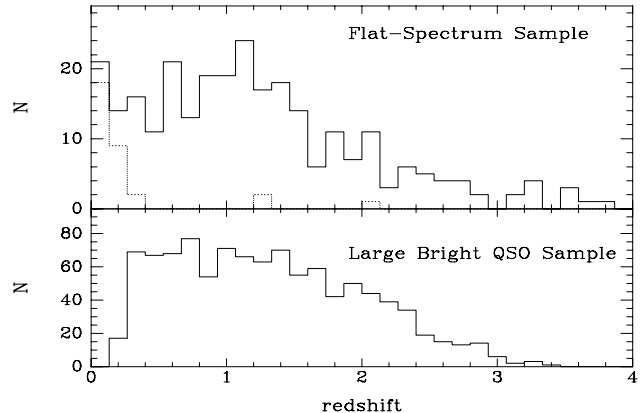}
\centering
\caption{Redshift histogram for the sample: the dotted line indicates
just the optically resolved sources (galaxies). The redshift histogram
of a large optically selected sample, the Large Bright QSO Survey is
shown in the lower panel for comparison.}
\label{fig_reds}
\end{figure}

Combining this new data with the published redshifts we now have
confirmed redshifts for 277 or 86\% of the sample and possible
redshifts for a further 10.  This represents a significant improvement
over the last major compilation of this sample (Savage et al.\ 1990)
when only 67\% of the redshifts were measured, and not all of them
published. A histogram of all the redshifts is given in
Fig.~\ref{fig_reds}.

\subsection{Notes on Individual Spectra}
\label{sec_snotes}

\begin{enumerate}

\item PKS~0114$+$074b: this is not part of the sample, but is close to
PKS~0114$+$074 and was the source of the previously quoted redshift
(see Section~\ref{sec_notes}).

\item PKS~0215$+$015: this is not part of the sample,
but was measured as part of a preliminary version of the sample and is
included here for reference.

\item PKS~1008$-$017: also observed with the AAT, 1988 May 11;
the combined spectrum was used.

\item PKS~1124$-$186: has weak lines but they were also observed on
the AAT 1984 May 01.

\item PKS~1336$-$260: also observed with the AAT, 1996 Apr 20;
combined spectrum used.

\item PKS~1557$+$032: this is not part of the sample,
but was measured as part of a preliminary version of the sample and is
included here for reference.

\item PKS~1648$+$015: also observed with the AAT, 1995 Jun 01; combined
spectrum used.

\item PKS~2021$-$330: possible broad absorption line structure near CIV.

\item PKS~2131$-$021: also observed with the ANU 2.3m Telescope, 1995 Sep
28; combined spectrum used. The redshift is based on OII and MgII in
our spectra and a reported ``definite'' line at 3541\AA\ (Baldwin et
al.\ 1989) which we identify with CIV.

\end{enumerate}

\section{The Catalogue}
\label{sec_cat}

We present all the data we have collected on our sample in
Table~\ref{tab_master}. We indicate the source of all published data
in the table by a reference number; the references are listed in
numerical order at the end of the paper. In all cases a minus sign in
front of the reference number indicates an uncertain value.  The
specific reference numbers 120 and 121 refer to new data we present in
this paper: 120 to accurate radio positions measured with the ATCA and
121 to all our other data including the VLA radio positions.

The columns in the table are as follows:
\begin{enumerate}

\item name: the Parkes source name.

\item $S_{2.7}, S_{5.0}, \alpha$, Rf: the 2.7 and 5.0\GHz\ source fluxes and
corresponding spectral index as published in reference Rf 
(see Table~\ref{radio_surveys}).

\item RA(B1950), Dec(B1950), Rc: the accurate B1950 
(i.e. equinox B1950 and epoch B1950)
radio source positions from reference Rc.

\item comment: (1) a brief description of the radio morphology if the
source is resolved using the terminology of Downes et al.\ (1986):
``P'' for partially resolved sources, ``Do'' for double sources with
no central component with the position defined by the centroid of the
source, ``Do+CC'' for double sources with a central component or peak
giving a well-defined position, ``H'' for a diffuse halo around a
central source, and ``HT'' for a complex head-tail morphology.  (2)
comments in parentheses refer to the optical identification.  In cases
where there was no match to the sky catalogues but the source was
identified using CCD data, these are indicated as ``(B)'', ``(R)'',
``(I)'' and ''(K)'' for the respective wavebands.  If the CCD imaging
did not identify the source, the comment ``null'' is made and
``STR'' indicates a source too near a bright star.  If the source was
confused with a close neighbour in the sky catalogues, but separated
by a CCD image the comment ``merge'' is made followed by the waveband
used; in some cases the Digitized Sky Survey data was used to separate
the object (``DSS'').

\item $\Delta RA, \Delta Dec, \Delta r$, cl, \bj: the position offsets
(arcsec, in the sense optical$-$radio) of the corresponding optical
image (if any), the total separation, the image classification (``g''
galaxy, ``s'' stellar, ``f'' too faint to classify, ``m'' merged) and
the apparent \bj\ magnitude if the counterpart was found in the sky
survey data.

\item z, Rz, Rsp: the emission redshift of the source obtained from
reference code Rz. If no emission redshift has been measured, but
absorption lines have been identified these are used to place a lower
limit on the source redshift indicated in the form ``$>$0.500''.  If a
spectrum has been published it can be found in reference Rsp.

\item RA(J2000), Dec(J2000): the corresponding 
J2000 positions.
\end{enumerate}

\section{Overview of the Sample}
\label{sec_corr}

We defer a detailed analysis of the sample to other papers, but we take
this opportunity to make a brief overview of the sample.

The sources with measured redshifts span the redshift range 0.05--3.78
with a median redshift of 1.07 (see Fig.~\ref{fig_reds}).  The
redshift histogram is smooth, and broadly similar to that of optical
surveys such as the Large Bright QSO Survey (LBQS; see Hewett et al.\
1995, and references therein).  The distribution of our sample is
compared to that of the LBQS in Fig.~\ref{fig_reds}. The lack of LBQS
quasars in the lowest redshift bin is due to the absolute magnitude
and redshift cut-off of that survey. The 2-sample Kolmogorov-Smirnov
test (comparing sources with redshift $z>0.22$ in both samples: 235
Parkes and 1018 LBQS) gives results consistent with the two samples
having the same redshift distribution at the 35\% probability level.
This makes the two samples ideal for comparing the properties of
radio- and optically-selected quasars.

\begin{figure}
\epsfxsize=\one_wide \epsffile{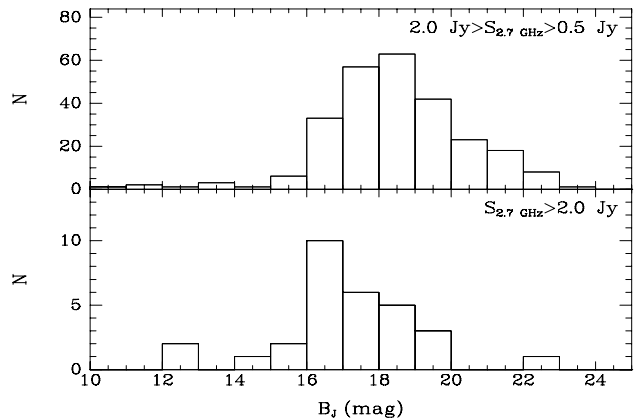}
 \centering
 \caption{The distribution of \bj\  magnitudes in the sample as a
function of 2.7\GHz\ radio fluxes.}
 \label{fig_mags2}
\end{figure}

The distribution of optical \bj\ magnitudes of our sources (see Figs.
\ref{fig_mags} and \ref{fig_mags2}) shows that despite the fact that
our sample was not selected on the basis of
optical magnitude, the sources occupy
a restricted range of magnitudes.  The majority have $\bj  = 18 \pm
3$. The well-defined mode in the distribution of \bj\  magnitudes is
not an artifact of the plate limit of $\bj  \approx 22.5$;
the number of  sources with $22 > \bj  > 20$ is clearly below that
with $20 > \bj  > 18$. However a small fraction of sources are clearly
very faint in \bj.

\begin{figure}
\epsfxsize=\one_wide \epsffile{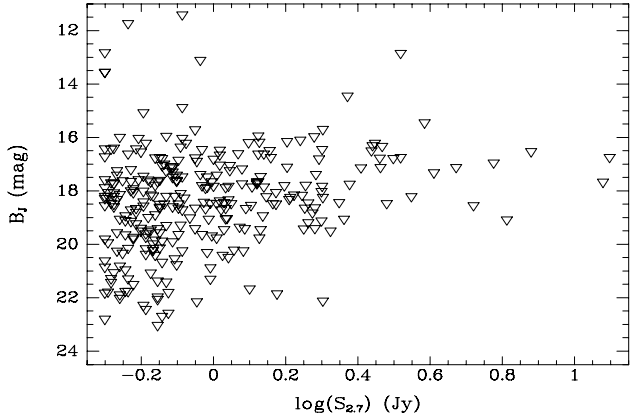}
 \centering
 \caption{The distribution of  \bj\  magnitudes as a function of 2.7\GHz\ 
radio flux for all sources in the survey.}
 \label{fig_opt_rad}
\end{figure}

In common with Browne \& Wright (1985) we find that the modal \bj\ 
magnitude of our flat-spectrum sample is a function of radio flux; the
most radio-bright sources have slightly brighter typical \bj\ 
magnitudes (Fig.~\ref{fig_mags2}). This is also shown in 
Fig.~\ref{fig_opt_rad}, a plot of the \bj\ magnitudes against the
2.7\GHz\ radio fluxes.

\begin{figure*}
\epsfxsize=\two_wide \epsffile{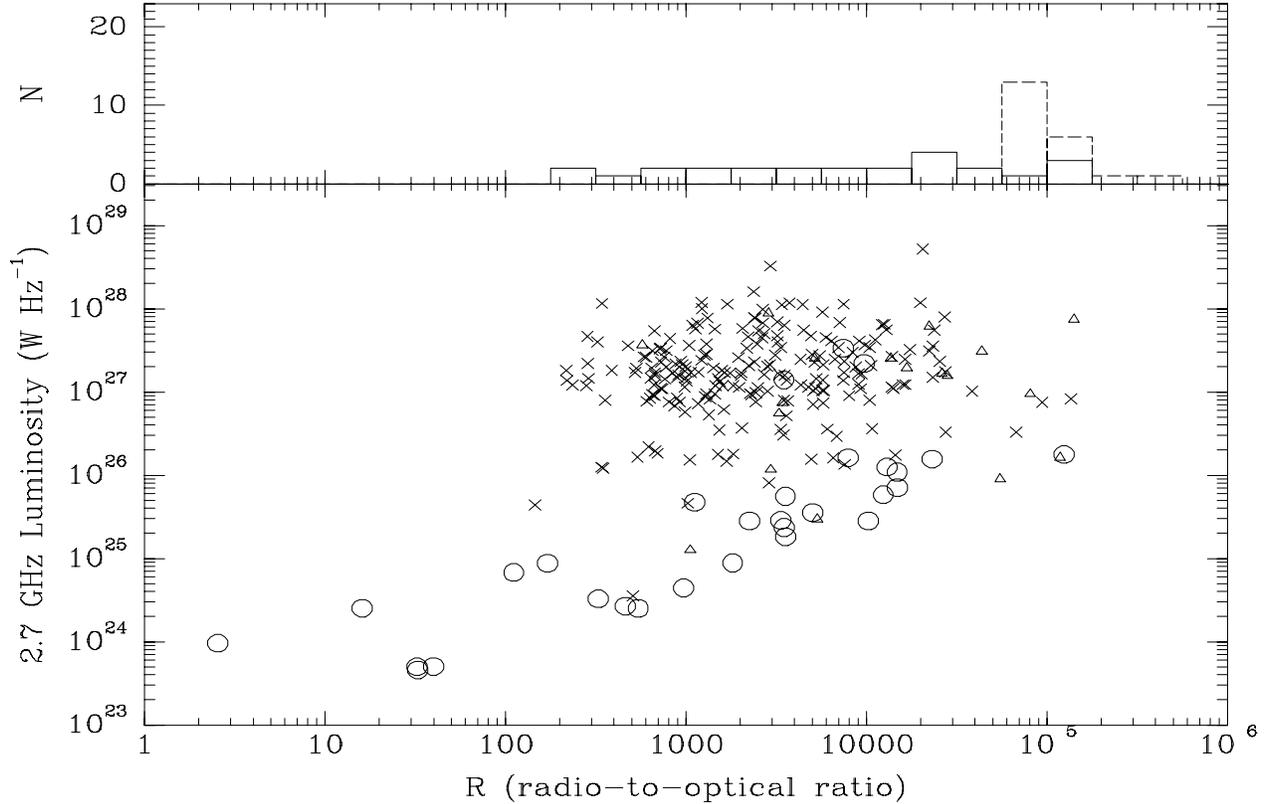}
 \centering

 \caption{Radio-to-optical ratios $R$ as a function of 2.7\GHz\
luminosity.  Absolute magnitudes and luminosities were computed
assuming $H_{\circ} = 75 {\thinspace\rm km\thinspace s}^{-1}{\rm
Mpc}^{-1}$ and $q_{\circ} = 0.5$; \bj\ magnitudes were $K$-corrected
assuming a continuum slope of $f_{\nu} \propto \nu^{-1}$, and radio
slope $f_{\nu} \propto \nu^{-0.09}$ which is the median slope of our
sources. Morphologically extended sources (as classified
automatically) are marked as circles and unresolved sources as crosses;
faint and merged sources are indicated by 
triangles. Radio-to-optical ratios of sources without measured
redshifts are presented as a histogram in the top panel; dotted lines
show lower limits on radio-to-optical ratios for sources not detected
on the sky surveys.
}
 \label{fig_ratio}
\end{figure*}

There is some suggestion from our data that the radio-to-optical ratio
may be a physically meaningful parameter, as originally suggested by
Schmidt (1970). In Fig.~\ref{fig_ratio} we plot radio-to-optical
ratios $R$ as a function of radio luminosity, for different classes of
source. A clear correlation can be seen, with the lowest
radio-luminosity sources having low values of $R$. If, however, we
exclude galaxies, the correlation disappears, and no sources remain
with $R<100$.
The sources without redshifts have radio-to-optical ratios above those
of the detected sources, which may reflect dust
obscuration of the \bj\  emission (Webster et al.\ 1995).

\begin{table*}
\vspace{-0.6cm}
\rotate{
\begin{tabular}{rrrrrrrrrlrrrrrlrrrr}
 N& name& $S_{2.7}$& $S_{5.0}$& $\alpha$& Rf& RA(B1950)& Dec(B1950)& Rc& comment& $\Delta$RA& $\Delta$Dec& $\Delta$r& cl& $B_J$& z& Rz& Rsp& RA(J2000)& Dec(J2000)\\
\\
     1& PKS0003$-$066&  1.46&  1.58&  0.13&  103
&   0:03:40.29&   $-$6:40:17.3&   56&               &  1.15
& $-$0.22&  1.17& s& 18.47& ~~0.347&   87&   87
&   0:06:13.89&   $-$6:23:35.2\\
     2& PKS0005$-$239&  0.58&  0.53& $-$0.15&  103
&   0:05:27.47&  $-$23:56:00.0&  121&             &  0.04
& $-$0.31&  0.31& s& 16.61& ~~1.407&  114&  108
&   0:08:00.37&  $-$23:39:18.0\\
     3& PKS0005$-$262&  0.58&  0.58&  0.00&  103
&   0:05:53.51&  $-$26:15:53.4&  121& (K)&  0.39
& $-$0.02&  0.39& s&  0.00& ~~0.0&    0&    0
&   0:08:26.25&  $-$25:59:11.5\\
     4& PKS0008$-$264&  0.67&  0.81&  0.31&  103
&   0:08:28.89&  $-$26:29:14.8&   57&               &  1.17
&  0.42&  1.24& s& 19.46& ~~1.096&  117&    0
&   0:11:01.25&  $-$26:12:33.3\\
     5& PKS0013$-$005&  0.89&  0.79& $-$0.19&  102
&   0:13:37.35&   $-$0:31:52.5&   57&               & $-$0.75
& $-$0.21&  0.78& s& 19.41& ~~1.574&  105&   25
&   0:16:11.08&   $-$0:15:12.3\\
     6& PKS0036$-$216&  0.53&  0.60&  0.20&  103
&   0:36:00.44&  $-$21:36:33.1&   57&               &  0.28
& $-$1.29&  1.31& g& 21.05& ~~0.0&    0&  121
&   0:38:29.95&  $-$21:20:03.9\\
\end{tabular}}
\caption{Master Source Catalogue}
\label{tab_master}
\end{table*}
\clearpage

\begin{table*}
\vspace{-0.6cm}
\rotate{
\begin{tabular}{rrrrrrrrrlrrrrrlrrrr}
 N& name& $S_{2.7}$& $S_{5.0}$& $\alpha$& Rf& RA(B1950)& Dec(B1950)& Rc& comment& $\Delta$RA& $\Delta$Dec& $\Delta$r& cl& $B_J$& z& Rz& Rsp& RA(J2000)& Dec(J2000)\\
\\
    41& PKS0229$-$398&  0.64&  0.68&  0.10&    6
&   2:29:51.99&  $-$39:49:00.2&   39& (mrg R)     & $-$0.69
&  0.86&  1.10& m& 22.28& ~~1.646& $-$121&  121
&   2:31:51.79&  $-$39:35:47.1\\
    42& PKS0232$-$042&  0.84&  0.62& $-$0.49&  103
&   2:32:36.51&   $-$4:15:08.9&   39&               &  0.62
& $-$1.15&  1.31& s& 16.21& ~~1.437&   84&   84
&   2:35:07.25&   $-$4:02:04.1\\
    43& PKS0237+040&  0.73&  0.77&  0.09&   79
&   2:37:14.41&    4:03:29.7&   56&               & $-$0.74
&  0.20&  0.77& s& 18.48& ~~0.978&   75&    0
&   2:39:51.26&    4:16:21.6\\
    44& PKS0238$-$084&  0.58&  1.40&  1.43&  103
&   2:38:37.36&   $-$8:28:09.0&   56&               & $-$0.34
&  0.53&  0.63& g& 11.73& ~~0.005&   67&    0
&   2:41:04.80&   $-$8:15:20.7\\
    45& PKS0240$-$217&  0.97&  0.82& $-$0.27&  103
&   2:40:19.23&  $-$21:45:11.6&  120&               &  0.99
&  1.01&  1.42& g& 19.05& ~~0.314&  117&  108
&   2:42:35.80&  $-$21:32:27.8\\
    46& PKS0240$-$060&  0.53&  0.52& $-$0.03&  103
&   2:40:43.24&   $-$6:03:37.5&  121&               &  0.18
& $-$0.86&  0.88& s& 18.20& ~~1.800&    3&    0
&   2:43:12.46&   $-$5:50:55.2\\
\end{tabular} 
}
\contcaption{Master Source Catalogue}
\end{table*}
\clearpage

\begin{table*}
\vspace{-0.6cm}
\rotate{
\begin{tabular}{rrrrrrrrrlrrrrrlrrrr}
 N& name& $S_{2.7}$& $S_{5.0}$& $\alpha$& Rf& RA(B1950)& Dec(B1950)& Rc& comment& $\Delta$RA& $\Delta$Dec& $\Delta$r& cl& $B_J$& z& Rz& Rsp& RA(J2000)& Dec(J2000)\\
\\
    81& PKS0434$-$188&  1.05&  1.19&  0.20&  103
&   4:34:48.97&  $-$18:50:48.2&   56&               & $-$0.46
&  0.06&  0.46& s& 18.70& ~~2.705&  107&  108
&   4:37:01.48&  $-$18:44:48.6\\
    82& PKS0438$-$436&  6.50&  7.00&  0.12&    6
&   4:38:43.18&  $-$43:38:53.1&   56&               & $-$0.05
& $-$0.21&  0.22& s& 19.08& ~~2.852&   52&   52
&   4:40:17.17&  $-$43:33:08.1\\
    83& PKS0440$-$003&  3.53&  3.13& $-$0.20&  102
&   4:40:05.29&   $-$0:23:20.6&   56&               & $-$0.31
& $-$0.60&  0.67& s& 18.21& ~~0.844&   75&    4
&   4:42:38.66&   $-$0:17:43.4\\
    84& PKS0445+097&  0.68&  0.56& $-$0.32&   79
&   4:45:36.99&    9:45:37.2&  121&               &  1.02
&  0.40&  1.10& g& 20.22& ~~2.115&   94&    5
&   4:48:21.70&    9:50:51.1\\
    85& PKS0448$-$392&  0.89&  0.89&  0.00&    6
&   4:48:00.45&  $-$39:16:15.7&   39&               &  0.51
& $-$0.18&  0.54& s& 16.76& ~~1.288&  114&  108
&   4:49:42.24&  $-$39:11:09.4\\
    86& PKS0451$-$282&  2.38&  2.50&  0.08&  103
&   4:51:15.13&  $-$28:12:29.3&   56&               &  0.84
&  0.05&  0.84& s& 17.75& ~~2.5637&    8&    8
&   4:53:14.64&  $-$28:07:37.2\\
\end{tabular}
}
\contcaption{Master Source Catalogue}
\end{table*}
\clearpage

\begin{table*}
\vspace{-0.6cm}
\rotate{
\begin{tabular}{rrrrrrrrrlrrrrrlrrrr}
 N& name& $S_{2.7}$& $S_{5.0}$& $\alpha$& Rf& RA(B1950)& Dec(B1950)& Rc& comment& $\Delta$RA& $\Delta$Dec& $\Delta$r& cl& $B_J$& z& Rz& Rsp& RA(J2000)& Dec(J2000)\\
\\
   121& PKS1021$-$006&  0.95&  0.75& $-$0.38&  102
&  10:21:56.19&   $-$0:37:41.6&  121&               & $-$0.09
&  0.23&  0.25& s& 17.90& ~~2.549&  121&  121
&  10:24:29.58&   $-$0:52:55.8\\
   122& PKS1032$-$199&  1.10&  1.15&  0.07&   70
&  10:32:37.37&  $-$19:56:02.2&   39&               & $-$0.45
&  0.15&  0.47& s& 18.33& ~~2.189&  107&  108
&  10:35:02.16&  $-$20:11:34.5\\
   123& PKS1034$-$293&  1.33&  1.51&  0.21&   70
&  10:34:55.83&  $-$29:18:27.0&   56&               & $-$0.62
&  0.76&  0.99& s& 15.94& ~~0.312&   87&  108
&  10:37:16.08&  $-$29:34:03.0\\
   124& PKS1036$-$154&  0.75&  0.78&  0.06&   70
&  10:36:39.48&  $-$15:25:28.1&   56&               & $-$0.54
&  0.42&  0.69& s& 21.80& ~~0.525&  121&  121
&  10:39:06.71&  $-$15:41:06.8\\
   125& PKS1038+064&  1.74&  1.40& $-$0.35&  $-$79
&  10:38:40.88&    6:25:58.3&  121& (mrgDSS& $-$0.15
& $-$0.22&  0.27& m& 16.10& ~~1.264&  121&  121
&  10:41:17.15&    6:10:16.6\\
   126& PKS1042+071&  0.50&  0.50&  0.00&   79
&  10:42:19.46&    7:11:25.0&  121&               & $-$1.03
&  0.00&  1.03& s& 18.55& ~~0.698&  105&    0
&  10:44:55.92&    6:55:37.9\\
\end{tabular} 
}
\contcaption{Master Source Catalogue}
\end{table*}
\clearpage

\begin{table*}
\vspace{-0.6cm}
\rotate{
\begin{tabular}{rrrrrrrrrlrrrrrlrrrr}
 N& name& $S_{2.7}$& $S_{5.0}$& $\alpha$& Rf& RA(B1950)& Dec(B1950)& Rc& comment& $\Delta$RA& $\Delta$Dec& $\Delta$r& cl& $B_J$& z& Rz& Rsp& RA(J2000)& Dec(J2000)\\
\\
   161& PKS1236+077&  0.59&  0.67&  0.21&   79
&  12:36:52.31&    7:46:45.4&   56&               & $-$0.46
& $-$0.80&  0.92& s& 19.14& ~~0.40&  105&  108
&  12:39:24.58&    7:30:17.1\\
   162& PKS1237$-$101&  1.35&  1.13& $-$0.29&    8
&  12:37:07.29&  $-$10:07:00.7&   57&               & $-$0.47
&  0.34&  0.58& s& 17.46& ~~0.751&  121&  121
&  12:39:43.07&  $-$10:23:28.9\\
   163& PKS1243$-$072&  0.79&  1.11&  0.55&    8
&  12:43:28.79&   $-$7:14:23.5&   57&               &  0.07
&  0.30&  0.31& s& 17.64& ~~1.286&  107&  108
&  12:46:04.23&   $-$7:30:46.7\\
   164& PKS1244$-$255&  1.34&  1.55&  0.24&   70
&  12:44:06.71&  $-$25:31:26.7&   57&               & $-$0.80
& $-$0.12&  0.81& s& 16.17& ~~0.638&  107&  108
&  12:46:46.79&  $-$25:47:49.4\\
   165& PKS1250$-$330&  0.52&  0.49& $-$0.10&   78
&  12:50:14.95&  $-$33:03:41.9&  121& P& $-$0.99
& $-$0.03&  0.99& s& 21.42& ~~0.0&    0&  121
&  12:52:58.47&  $-$33:19:59.0\\
   166& PKS1253$-$055& 12.00& 13.00&  0.13&    8
&  12:53:35.84&   $-$5:31:08.0&   56&               & $-$0.32
& $-$0.07&  0.33& s& 17.67& ~~0.540&   13&    0
&  12:56:11.17&   $-$5:47:21.7\\
\end{tabular} 
}
\contcaption{Master Source Catalogue}
\end{table*}
\clearpage

\begin{table*}
\vspace{-0.6cm}
\rotate{
\begin{tabular}{rrrrrrrrrlrrrrrlrrrr}
 N& name& $S_{2.7}$& $S_{5.0}$& $\alpha$& Rf& RA(B1950)& Dec(B1950)& Rc& comment& $\Delta$RA& $\Delta$Dec& $\Delta$r& cl& $B_J$& z& Rz& Rsp& RA(J2000)& Dec(J2000)\\
\\
   201& PKS1437$-$153&  0.72&  0.64& $-$0.19&   70
&  14:37:11.35&  $-$15:18:58.8&   57&               & $-$1.24
& $-$0.02&  1.24& s& 19.87& ~~0.0&    0&    0
&  14:39:56.88&  $-$15:31:50.6\\
   202& PKS1438$-$347&  0.50&  0.45& $-$0.17&   78
&  14:38:20.36&  $-$34:43:57.5&  121&               & $-$2.30
& $-$0.14&  2.30& s& 17.58& ~~1.159&   37&    0
&  14:41:24.01&  $-$34:56:45.8\\
   203& PKS1443$-$162&  0.78&  0.65& $-$0.30&   70
&  14:43:06.68&  $-$16:16:26.7&   57&               & $-$0.70
&  0.04&  0.70& s& 20.53& ~~0.0&    0&    0
&  14:45:53.37&  $-$16:29:01.7\\
   204& PKS1445$-$161&  1.06&  0.80& $-$0.46&   70
&  14:45:28.34&  $-$16:07:56.5&  120&               & $-$0.68
&  0.49&  0.84& s& 20.40& ~~2.417&  121&  121
&  14:48:15.06&  $-$16:20:24.7\\
   205& PKS1450$-$338&  0.72&  0.54& $-$0.47&   78
&  14:50:58.10&  $-$33:48:46.0&  121& (mrg K)& $-$1.01
&  0.07&  1.01& f&  0.00& ~~0.368&  121&  121
&  14:54:02.59&  $-$34:00:57.6\\
   206& PKS1454$-$060&  0.83&  0.62& $-$0.47&    8
&  14:54:02.68&   $-$6:05:39.1&  121&               & $-$1.84
& $-$1.68&  2.49& s& 18.27& ~~1.249&   12&    0
&  14:56:41.48&   $-$6:17:42.0\\
\end{tabular} 
}
\contcaption{Master Source Catalogue}
\end{table*}
\clearpage

\begin{table*}
\vspace{-0.6cm}
\rotate{
\begin{tabular}{rrrrrrrrrlrrrrrlrrrr}
 N& name& $S_{2.7}$& $S_{5.0}$& $\alpha$& Rf& RA(B1950)& Dec(B1950)& Rc& comment& $\Delta$RA& $\Delta$Dec& $\Delta$r& cl& $B_J$& z& Rz& Rsp& RA(J2000)& Dec(J2000)\\
\\
   241& PKS1933$-$400&  1.20&  1.44&  0.30&    6
&  19:33:51.12&  $-$40:04:46.8&   56&               &  0.41
& $-$1.28&  1.34& s& 17.73& ~~0.965&  121&  121
&  19:37:16.21&  $-$39:58:00.8\\
   242& PKS1953$-$325&  0.51&  0.63&  0.34&   78
&  19:53:48.42&  $-$32:33:49.9&   57&               &  0.36
&  1.71&  1.74& s& 19.94& ~~1.242&   37&  108
&  19:56:59.49&  $-$32:25:46.2\\
   243& PKS1954$-$388&  2.00&  2.00&  0.00&    6
&  19:54:39.06&  $-$38:53:13.3&   56&               & $-$0.54
&  0.20&  0.58& s& 17.82& ~~0.626&  107&  108
&  19:57:59.83&  $-$38:45:06.1\\
   244& PKS1958$-$179&  1.11&  1.17&  0.09&    7
&  19:58:04.61&  $-$17:57:16.9&   56&               &  0.30
& $-$0.02&  0.30& s& 17.05& ~~0.652&  121&  121
&  20:00:57.09&  $-$17:48:57.5\\
   245& PKS2000$-$330&  0.71&  1.20&  0.85&   78
&  20:00:13.02&  $-$33:00:12.5&   56& (mrgDSS&  0.13
&  0.03&  0.13& m& 19.56& ~~3.7832&   94&    5
&  20:03:24.12&  $-$32:51:44.4\\
   246& PKS2002$-$185&  0.64&  0.48& $-$0.47&    7
&  20:02:24.43&  $-$18:30:39.0&  121&               & $-$0.10
& $-$0.45&  0.46& s& 17.44& ~~0.859&  107&  108
&  20:05:17.32&  $-$18:22:03.3\\
\end{tabular}
}
\contcaption{Master Source Catalogue}
\end{table*}
\clearpage

\begin{table*}
\vspace{-0.6cm}
\rotate{
\begin{tabular}{rrrrrrrrrlrrrrrlrrrr}
 N& name& $S_{2.7}$& $S_{5.0}$& $\alpha$& Rf& RA(B1950)& Dec(B1950)& Rc& comment& $\Delta$RA& $\Delta$Dec& $\Delta$r& cl& $B_J$& z& Rz& Rsp& RA(J2000)& Dec(J2000)\\
\\
   281& PKS2210$-$257&  0.96&  1.02&  0.10&  103
&  22:10:14.13&  $-$25:44:22.5&   56&               & $-$1.70
& $-$1.12&  2.03& s& 17.86& ~~1.833&  107&  108
&  22:13:02.50&  $-$25:29:30.1\\
   282& PKS2212$-$299&  0.54&  0.44& $-$0.33&  103
&  22:12:25.11&  $-$29:59:19.8&  121&               & $-$0.44
& $-$0.24&  0.50& s& 17.24& ~~2.703&    5&    5
&  22:15:16.03&  $-$29:44:23.0\\
   283& PKS2215+020&  0.70&  0.64& $-$0.15&  102
&  22:15:15.59&    2:05:09.0&   57&               & $-$0.74
& $-$0.50&  0.89& s& 21.97& ~~3.572&  121&  121
&  22:17:48.24&    2:20:10.9\\
   284& PKS2216$-$038&  1.04&  1.30&  0.36&  102
&  22:16:16.38&   $-$3:50:40.7&   56&               & $-$0.23
& $-$0.38&  0.45& s& 16.65& ~~0.901&  109&  109
&  22:18:52.03&   $-$3:35:36.8\\
   285& PKS2223$-$052&  4.70&  4.31& $-$0.14&  103
&  22:23:11.08&   $-$5:12:17.8&   57&               &  0.64
& $-$0.30&  0.71& s& 17.12& ~~1.404&   73&    0
&  22:25:47.26&   $-$4:57:01.3\\
   286& PKS2227$-$088&  1.49&  1.41& $-$0.09&  103
&  22:27:02.34&   $-$8:48:17.6&   56&               &  1.20
&  0.02&  1.20& s& 18.05& ~~1.561&  107&  108
&  22:29:40.08&   $-$8:32:54.3\\
\end{tabular}}
\contcaption{Master Source Catalogue}
\end{table*}
\clearpage

This correlation can be modelled by assuming a strict proportionality
between the \bj\ and 2.7\GHz\ luminosities of our quasars, but adding
the light of a host galaxy. If we assume that all host galaxies have
absolute \bj\  magnitudes of $\sim -20.5$, the quasar light will
dominate over the host galaxy light for 2.7\GHz\ luminosities $>10^{26}
{\rm W \ Hz}^{-1}$ (for assumed cosmology see the caption to
Fig.~\ref{fig_ratio}). Above this luminosity there will be no
correlation between $R$ and the radio flux, as observed, and below this
luminosity $R$ will be proportional to the radio flux, which
is consistent with the observed correlation.

The radio spectral indices do not correlate with redshift, apparent 
\bj\  magnitude, radio flux or radio luminosity---but the range of 
spectral index in the sample is of course limited.

\section*{Acknowledgements}

The compilation of this paper would not have been possible without the
efforts of many people who have worked on the Parkes radio samples over
the past 20 years or more. We particularly acknowledge the work done by
Graeme White on the optical identifications, Saul Caganoff on the 
early stages of our image analysis, and Alan Wright and Robina
Otrupcek in compiling the machine-readable version of the Parkes Catalogue.

Our new radio observations were made with the VLA and the ATCA. John
Reynolds and Lucyna Kedziora-Chudczer assisted with the ATCA
observations and we would also like to thank John Reynolds for helpful
comments about astrometry.

We would like to thank Raylee Stathakis for making our service
observations on the AAT, Russell Cannon, Director of the AAO for
awarding us additional discretionary observing time and Roy Antaw for
extracting large amounts of data from the AAT archive for us.  We are
also grateful to Katrina Sealey for obtaining some additional data for
us with the ANU 2.3m Telescope and Mike Bessell for giving technical
advice about the 2.3m out-of-hours.

We have made substantial use of the following on-line databases in the
compilation of this paper: The APM Catalogues (we thank Mike Irwin for
scanning additional fields for us); the Center for Astrophysics
Redshift Survey (with kind assistance from Cathy Clemens); the
COSMOS/UKST Southern Sky Catalogue supplied by the Anglo-Australian
Observatory; the Lyon-Meudon Extragalactic Database (LEDA) supplied by
the LEDA team at the CRAL-Observatoire de Lyon (France); the NASA/IPAC
Extragalactic Database (NED) which is operated by the Jet Propulsion
Laboratory, Caltech, under contract with the National Aeronautics and
Space Administration.

The Digitized Sky Survey was produced at the Space Telescope Science
Institute under U.S. Government grant
NAG W-2166. The images are based on photographic data
obtained using the Oschin Schmidt
Telescope on Palomar Mountain and the UK Schmidt Telescope. The plates were
processed into 
compressed digital form with the permission of these institutions. 

The Palomar Observatory Sky Survey was funded by the National Geographic
Society. The Oschin Schmidt
Telescope is operated by the California Institute of Technology and Palomar
Observatory. 

The UK Schmidt Telescope was operated by the Royal Observatory Edinburgh,
with funding from the UK Science
and Engineering Research Council, until 1988 June, and thereafter by the
Anglo-Australian Observatory. Original
plate material is copyright the Royal Observatory Edinburgh and the
Anglo-Australian Observatory.

Finally we wish to thank the referee for many helpful suggestions.

\section*{References}

This section lists the references in alphabetical order, each followed
by a code number used to refer to the reference in Table~\ref{tab_master}.

\smallskip

\refs Baars, J. W. M., Genzel, R., Pauliny-Toth, I. I. K., Witzel, A., 1977, A\&A, 61, 99 (001) 
\refs Baldwin, J. A., 1975, ApJ, 201, 26 (002) 
\refs Baldwin, J. A., Wampler, E. J., Burbidge, E. M., 1981, ApJ, 243, 76 (003) 
\refs Baldwin, J. A., Wampler, E. J., Gaskell, C. M., 1989, ApJ, 338, 630 (004) 
\refs Barthel, P. D., Tytler, D. R., Thomson, B., 1990, A\&AS, 82, 339 (005) 
\refs Bolton, J. G., Shimmins, A. J., 1973, Aust. J. Phys., Astrophys. Suppl., 30, 1 (006) 
\refs Bolton, J. G., Shimmins, A. J., Wall, J. V., 1975, Aust. J. Phys., Astrophys. Suppl., 34, 1 (007) 
\refs Bolton, J. G., Savage, A., Wright, A. E., 1979, Aust. J. Phys., Astrophys. Suppl., 46, 1 (008) 
\refs Browne, I. W. A., Wright, A. E., 1985, MNRAS, 213, 97 (009) 
\refs Browne, I. W. A., Savage, A., Bolton, J. G., 1975, MNRAS, 173, 87p (010) 
\refs Burbidge, E. M., 1967, ApJ, 149, L51 (011) 
\refs Burbidge, E. M., 1968, ApJ, 154, L109 (012) 
\refs Burbidge, E. M., Rosenberg, F. D., 1965, ApJ, 142, 1673 (013) 
\refs Burbidge, E. M., Strittmatter, P. A., 1972, ApJ, 174, L57 (014) 
\refs Caganoff, S., 1989, Ph. D. Thesis. Australian National University, Canberra (015) 
\refs Condon, J. J., Broderick, J. J., Seielstad, G. A., 1991, AJ, 102, 2041 (016) 
\refs Condon, J. J., Hicks, P. D., Jauncey, D. L., 1977, AJ, 82, 692 (017) 
\refs Da Costa, L. N., Nunes, M. A., Pellegrini, P. S., Willmer, C., Chincarini, G., Cowen, J. J., 1986, AJ, 91, 6 (018) 
\refs Danziger, I. J., Goss, W. M., 1983, MNRAS, 202, 703 (019) 
\refs Dekker, H., D'Odorico, S., 1984, Messenger, no. 37, 7 (020) 
\refs de Vaucouleurs, G., de Vaucouleurs, A., Corwin Jr., H. G., Buta, R. J., Paturel, G., Fouqu\'e, P., 1991, Third Reference Catalogue of Bright Galaxies. Springer-Verlag, New York (021) 
\refs Downes, A. J. B., Peacock, J. A., Savage, A., Carrie, D. R., 1986, MNRAS, 218, 31 (022) 
\refs Drinkwater, M. J., Schmidt, R. W., 1996, PASA, 13, 127 (023) 
\refs Drinkwater, M.J., Barnes, D.G., Ellison, S.L., 1995, PASA, 12, 248 (024) 
\refs Dunlop, J. S., Peacock, J. A., Savage, A., Lilly, S. J., Heasley, J. N., Simon, A. J. B., 1989, MNRAS, 238, 1171 (025) 
\refs Fairall, A. P. et al., 1992, AJ, 103, 11 (026) 
\refs Fiedler, R. L. et al., 1987, ApJS, 65, 319 (027) 
\refs Foltz, C. B., Chaffee, F. H., Hewett, P. C., Weymann, R. J., Anderson, S. F., MacAlpine, G. M., 1989, AJ, 98, 1959 (028) 
\refs Helou, G., Madore, B. F., Schmitz, M., Bicay, M. D., Wu, X., Bennett, J., 1991, in Egret, D., Albrecht, M., eds, Databases and On-Line Data in Astronomy. Kluwer, Dordrecht, p. 89 (029) 
\refs Hewett, P. C., Foltz, C. B., Chaffee, F. H., Francis, P. J., Weymann, R. J., Morris, S. L., Anderson, S. F., MacAlpine, G. M., 1991, AJ, 101, 1121 (030) 
\refs Hewett, P. C., Foltz, C. B., Chaffee, F. H., 1995, AJ, 109, 1498 (031) 
\refs Hewitt, A., Burbidge, G., 1993, ApJS, 87, 451 (032) 
\refs Huchra, J., Geller, M., Clemens, C., Tokarz, S., Michel, A., 1992, Bull. Inf. Cent. Donn\'ees Astron. Strasb. 41, 31 (033) 
\refs Hunstead, R. W., Murdoch, H. S., Shobbrook, R. R., 1978, MNRAS, 185, 149 (034) 
\refs Irwin, M., Maddox, S., McMahon, R., 1994, Spectrum: Newsletter of the Royal Observatories, no. 4, 14 (035) 
\refs Jauncey, D. L., Wright, A. E., Peterson, B. A., Condon, J. J., 1978, ApJ, 219, L1 (036) 
\refs Jauncey, D. L., Batty, M. J., Gulkis, S., Savage, A., 1982, AJ, 87, 763 (037) 
\refs Jauncey, D. L., Batty, M. J., Wright, A. E., Peterson, B. A., Savage, A., 1984, ApJ, 286, 498 (038) 
\refs Jauncey, D. L., Savage, A, Morabito, D. D., Preston, R. A., Nicholson, G. D., Tzioumis, A. K., 1989, AJ, 98, 54 (039) 
\refs Johnston, K. J. et al., 1995, AJ, 110, 880 (040) 
\refs K\"uhr, H., Witzel, A., Pauliny-Toth, I. I. K., Nauber, U., 1981, A\&AS, 45, 367 (041) 
\refs Laing, R. A., Riley, J. M., Longair, M. S., 1983, MNRAS, 204, 151 (042) 
\refs Lister, M. L., Gower, A. C., Hutchings, J. B., 1994, AJ, 108, 821 (043) 
\refs Lynds, C. R., 1967, ApJ, 147, 837 (044) 
\refs Ma, C., Shaffer, D. B., De Vegt, C., Johnston, K. J., Russell, J. L., 1990, AJ, 99, 1284 (045) 
\refs Maoz, D. et al., 1993, ApJ, 409, 28 (046) 
\refs Maza, J., Ruiz, M.-T., 1989, ApJS, 69, 353, (047) 
\refs Melnick, J., Quintana, H., 1981, AJ, 86, 1567 (048) 
\refs Metcalfe, N., Fong, R., Shanks, T., Kilkenny, D., 1989, MNRAS, 236, 207 (049) 
\refs Morabito, D. D., Preston, R. A., Slade, M. A., Jauncey, D. L., 1982, AJ, 87, 517 (050) 
\refs Morris, S. L., Ward, M. J., 1988, MNRAS, 230, 639 (051) 
\refs Morton, D. C., Savage, A., Bolton, J. G., 1978, MNRAS, 185, 735 (052) 
\refs Murdoch, H. S., Hunstead, R. W., White, G. L., 1984., PASA, 5, 341 (053) 
\refs Oke, J. B., Shields, G. A., Korycansky, D. G., 1984, ApJ, 277, 64 (054) 
\refs Patnaik, A., 1996, in preparation (055) 
\refs Perley, R. A., 1982, AJ, 87, 859 (056) 
\refs Perley, R.A., Taylor, G.B., 1996, The VLA Calibrator Manual. National Radio Astronomy Observatory, Socorro (057) 
\refs Peterson, B. A., Jauncey, D. L., Wright, A. E., Condon, J. J., 1976, ApJ, 207, L5 (058) 
\refs Peterson, B. A., Wright, A. E., Jauncey, D. L., Condon, J. J., 1979, ApJ, 232, 400 (059) 
\refs Preston, R. A., Morabito, D. D., Williams, J. G., Faulkner, J., Jauncey, D. L., Nicolson, G. D. 1985, AJ, 90, 1599 (060) 
\refs R\"oser, S., Bastian, U., Kuzmin, A, 1994, A\&AS, 105, 301 (061) 
\refs Reynolds, J. E. et al., 1995, A\&A, 304, 116 (062) 
\refs Richstone, D. O., Schmidt, M., 1980, ApJ, 235, 361 (063) 
\refs Russell, J. L. et al., 1994, Astron. J., 107, 379 (064) 
\refs Sandage, A., 1966, ApJ, 145, 1 (065) 
\refs Sargent, W. L. W., 1970, ApJ, 160, 405 (066) 
\refs Sargent, W. L. W., Schechter, P. L., Boksenberg, A., Shortridge, K., 1977, ApJ, 212, 326 (067) 
\refs Sargent, W. L. W., Steidel, C. C., Boksenberg, A., 1989, ApJS, 69, 703 (068) 
\refs Savage, A., Browne, I. W. A., Bolton, J. G., 1976, MNRAS, 177, 77p (069) 
\refs Savage, A., Wright, A. E., Bolton, J. G., 1977, Aust. J. Phys., Astrophys. Suppl., 44, 1 (070) 
\refs Savage, A., Clowes, R. G., Cannon, R. D., Cheung, K., Smith, M. G., Boksenberg, A., Wall, J. V., 1985, MNRAS, 213, 485 (071) 
\refs Savage, A., Jauncey, D. L., White, G. L., Peterson, B. A., Peters W. L., Gulkis, S., Condon, J. J., 1990, Aust. J. Phys., 43, 241 (072) 
\refs Schmidt, M., 1966, ApJ, 144, 443 (073) 
\refs Schmidt, M., 1970, ApJ, 162, 371 (074) 
\refs Schmidt, M., 1977, ApJ, 217, 358 (075) 
\refs Schmidt, M., Green, R. F., 1983, ApJ, 269, 352 (076) 
\refs Shepherd, M. C., Pearson, T. J., Taylor, G. B., 1995, BAAS, 27, 903 (077) 
\refs Shimmins, A. J., Bolton, J. G., 1974, Aust. J. Phys., Astrophys. Suppl., 32, 1 (078) 
\refs Shimmins, A. J., Bolton, J. G., Wall, J. V., 1975, Aust. J. Phys., Astrophys. Suppl., 34, 63 (079) 
\refs Smith, H. E., Jura, M., Margon, B., 1979, ApJ, 228, 369 (080) 
\refs Spinrad, H., Liebert, J., Smith, H. E., Hunstead, R. W., 1976, ApJ, 206, L79 (081) 
\refs Spinrad, H., Marr, J., Aguilar, L., Djorgovski, S., 1985, PASP, 97, 932 (082) 
\refs Stannard, D., Bentley, M., 1977, MNRAS, 180, 703 (083) 
\refs Steidel, C. C., Sargent, W. L. W., 1991, ApJ, 382, 433 (084) 
\refs Steidel, C. C., Sargent, W. L. W., 1992, ApJS, 80, 1 (085) 
\refs Stickel, M., K\"uhr, H., 1993, A\&AS, 100, 395 (086) 
\refs Stickel, M., Fried, J. W., K\"uhr, H., 1989, A\&AS, 80, 103 (087) 
\refs Stickel, M., Padovani, P., Urry, C. M., Fried, J. W., K\"uhr, H., 1991, ApJ, 374, 431 (088) 
\refs Stickel, M., K\"uhr, H., Fried, J. W., 1993, A\&AS, 97, 483 (089) 
\refs Stickel, M., Fried, J. W., K\"uhr, H., 1993, A\&AS, 98, 393 (090) 
\refs Stickel, M., Meisenheimer, K., K\"uhr, H., 1994, A\&AS, 105, 211 (091) 
\refs Sutherland, W., Saunders, W., 1992, MNRAS, 259, 413 (092) 
\refs Tadhunter, C. N., Morganti, R., di Serego Alighieri, S., Fosbury, R. A. E., Danziger, I. J., 1993, MNRAS, 263, 999 (093) 
\refs Tytler, D., Fan, X.-M., 1992, ApJS, 79, 1 (094) 
\refs Tytler, D., Boksenberg, A., Sargent, W. L. W., Young, P., Kunth, D., 1987, ApJS, 64, 667 (095) 
\refs Ulrich, M.-H., 1981, A\&A, 103, L1 (096) 
\refs Ulvestad, J., Johnston, K., Perley, R., Fomalont, E., 1981, AJ, 86, 1010 (097) 
\refs V\'eron, P., V\'eron-Cetty, M.-P., Djorgovski, S., Magain, P., Meylan, G., Surdej, J., 1990, A\&AS, 86, 543 (098) 
\refs V\'eron-Cetty, M.-P., V\'eron, P, 1993, ESO Sci. Rep., 13, 1 (099) 
\refs Wall, J. V., 1972, Aust. J. Phys., Astrophys. Suppl., 24, 1 (100) 
\refs Wall, J. V., Peacock, J. A., 1985, MNRAS, 216, 173 (101) 
\refs Wall, J. V., Shimmins, A. J., Merkelijn, J. K., 1971, Aust. J. Phys., Astrophys. Suppl., 19, 1 (102) 
\refs Wall, J. V., Wright, A. E., Bolton, J. G., 1976, Aust. J. Phys., Astrophys. Suppl., 39, 1 (103) 
\refs Webster, R. L., Francis, P. J., Peterson, B. A., Drinkwater, M. J., Masci, F. J., 1995, Nature, 375, 469 (104) 
\refs White, G. L., Jauncey, D. L., Savage, A., Wright, A. E., Batty, M. J., Peterson, B. A., Gulkis, S., 1988, ApJ, 327, 561 (105) 
\refs Wilkes, B. J., 1984, MNRAS, 207, 73 (106) 
\refs Wilkes, B. J., 1986, MNRAS, 218, 331 (107) 
\refs Wilkes, B. J., Wright, A. E., Jauncey, D. L., Peterson, B. A., 1983, PASA, 5, 2 (108) 
\refs Wills, B. J., Netzer, H., Uomoto, A. K., Wills, D., 1980, ApJ, 237, 319 (109) 
\refs Wills, D., Lynds, R., 1978, ApJS, 36, 317 (110) 
\refs Wills, D., Wills, B. J., 1974, ApJ, 190, 271 (111) 
\refs Wills, D., Wills, B. J., 1976, ApJS, 31, 143 (112) 
\refs Wright, A. E., Otrupcek, R. E., 1990, Parkes Catalogue. Australia Telescope National Facility, Epping (PKSCAT90) (113) 
\refs Wright, A. E., Jauncey, D. L., Peterson, B. A., Condon, J. J., 1977, ApJ, 211, L115 (114) 
\refs Wright, A. E., Peterson, B. A., Jauncey, D. L., Condon, J. J., 1978, ApJ, 226, L61 (115) 
\refs Wright, A. E., Peterson, B. A., Jauncey, D. L., Condon, J. J., 1979, ApJ, 229, 73 (116) 
\refs Wright, A. E., Ables, J. G., Allen, D. A., 1983, MNRAS, 205, 793 (117) 
\refs Yentis, D.J., Cruddace, R.G., Gursky, H., Stuart, B.V., Wallin, J.F., MacGillivray, H.T., Collins, C.A., 1992, in MacGillivray, H.T., Thomson, E.B., eds, Digitised Optical Sky Surveys. Kluwer, Dordrecht, p. 67 (118) 
\refs Young, P., Sargent, W. L. W., Boksenberg, A., 1982, ApJS, 48, 455 (119) 
\refs this paper (ATCA radio data) (120) 
\refs this paper (all optical and VLA radio data) (121) 

\newpage

\noindent Please note the following sections are not included in this
preprint but may be obtained from my
preprint page 
http://www.aao.gov.au/local/www/mjd/papers/.

\appendix

\section{Radio Images of Resolved Sources}
\label{ap_map}

{\bf Figure A1:} Radio images of resolved sources from the sample. 
There is one negative contour (dotted) shown in each image
with the same absolute value but opposite sign as the lowest positive
contour. The remaining positive contours are spaced logarithmically,
each level a factor of 2 larger than the previous one. The title of
each image indicates the level of the lowest positive contour as well
as the telescope used: the VLA images are at 4.885\GHz\ and the ATCA
images are at 4.796\GHz.

\section{Optical Finding Charts of All Sources}
\label{sec_charts}

{\bf Figure B1:} Finding charts for all the sources generated
from the automated sky catalogues.  Images classified automatically as
unresolved (stellar) are plotted as filled ellipses on the charts;
resolved images (galaxies) are plotted as unfilled ellipses.  The
charts are a good approximation to the photographic data, but we
stress that there can be problems with image merging in crowded
fields: close objects (e.g. two stars) can be misclassified as a
``merged'' object or galaxy. The ``Field'' code at the bottom of each
chart indicates the UKST field number (or the plate number for POSS~I)
with a prefix describing the type of plate. The prefix ``J'' indicates
UKST \Bj\ plates measured by COSMOS. For APM data ``j'' indicates UKST
\Bj\ plates, ``O'' blue POSS~I plates and ``E'' red POSS~I plates.

\section{New Optical Spectra}
\label{sec_spectra}

(to go on microfiche)

\end{document}